\documentclass[12pt]{article}
\usepackage{amssymb}
\usepackage{amsmath}
\usepackage{appendix}
\usepackage{graphicx}
\usepackage{color}

\title{Vanishing Poynting observers and electromagnetic field classification in Kerr and Kerr-Newman spacetimes}

\author{ H. Vargas--Rodr\'iguez$^1$, H. C. Rosu$^2$,  M. G. Medina--Guevara$^1$,\\  A. Gallegos$^1$, M.A. Mu\~niz--Torres$^1$}

\begin{document}
\maketitle

{\small
\noindent
$^1$Departamento de Ciencias Exactas y Tecnolog\'{\i}a, Centro Universitario de los Lagos, Universidad de Guadalajara,
  Enrique D\'{\i}az de Le\'on 1144, Col. Paseos de la Monta\~na, Lagos de Moreno, Jalisco, Mexico\\
$^2$IPICyT, Instituto Potosino de Investigacion Cientifica y Tecnologica,
Camino a la presa San Jos\'e 2055, Col. Lomas 4a Secci\'on, 78216 San Luis Potos\'{\i}, S.L.P., Mexico


\begin{abstract}
\noindent
We consider electromagnetic fields having an angular momentum density in a locally non-rotating reference frame in
Schwarzschild, Kerr, and Kerr-Newman spacetimes. The nature of such fields is assessed with two families of
observers, the locally non-rotating ones and those of vanishing Poynting flux. The velocity fields
of the vanishing-Poynting observers in the locally non-rotating reference frames are determined using the 3+1 decomposition formalism.
From a methodological point of view, and considering a classification of the electromagnetic field based on its
invariants, it is convenient to separate the consideration of the vanishing-Poynting observers into two cases corresponding to the pure and non-pure
fields, additionally if there are regions where the field rotates with the speed of light (light surfaces) it becomes necessary to split these
observers into two subfamilies. We present several examples of relevance in astrophysics and general relativity, such as pure rotating dipolar-like
magnetic fields and the electromagnetic field of the Kerr-Newman solution. For the latter example, we see that vanishing-Poynting observers
also measure a vanishing super-Poynting vector, confirming recent results in the literature. Finally, for all non-null electromagnetic fields, we present
the 4-velocity fields of vanishing Poynting observers in an arbitrary spacetime.

\end{abstract}





\section{Introduction}

In General Relativity, test electromagnetic fields are those that do not perturb the spacetime geometry.
In the case of magnetic test fields around a black hole of mass $M$, they should satisfy \cite{Kolos}
\begin{equation}
B<<\frac{c^4}{G^{3/2}}M_\odot\left( \frac{M_\odot}{M}\right)\sim \left( \frac{M_\odot}{M}\right)10^{19}\,{\rm Gauss}~,
\end{equation}
where $M_\odot$ is the mass of the sun.
According to  estimations given in \cite{piotrovich2010magnetic},
typical magnetic field scales are of the order $B_1\sim 10^8$G near the horizon of a stellar
mass black hole ($M\sim 10M_\odot$), and of the order $B_2\sim 10^4$G near the horizon
of a supermassive  black hole ($M\sim  10^9 M_\odot$).
For the supermassive black hole in M87, recent observations with the Event Horizon Telescope have led to an estimated
magnetic field strength of only 1 - 30 G
near the event horizon \cite{Akiyama}.
Hence all the above estimated strengths are clearly weak test fields. Even the magnetic
field $B\sim10^{15}$G of a ultramagnetized neutron star ($M<3M_\odot$)
could be considered as a test one \cite{Ibrahim}. So, it appears that important astrophysical electromagnetic fields
can be regarded as test ones.

In particular, electromagnetic fields on different gravitational backgrounds have been important to understand the nature and
dynamics of accretion disks and relativistic jets \cite{Kolos, Tursunov, Kolos2017, Mustafa, Frolov, Igata,  Bacchini, Zamaninasab},
pulsar magnetospheres \cite{Kalapotharakos, Petri1, Cerutti}, black hole magnetospheres \cite{ Palenzuela, Nathanail, Levinson},
or to study the propagation of electromagnetic waves in the neighborhood of massive bodies \cite{Asenjo}.

In this paper, we extend the analysis in our previous work \cite{Nuestroarticulo}, where we focused on rotating electromagnetic
fields in Minkowski space times. Since these fields have a Poynting vector tangent to closed lines
 around the rotation axis, they have an angular momentum density \cite{Jackson} that can be also
determined using the Noether's theorem \cite{Nova}.
To describe this class of electromagnetic fields, several families of observers were introduced, among which
a special family of rotating observers for which the Poynting vector vanishes and as such they measure no net flux of electromagnetic energy,
together with the opposite family of inertial observers at rest with respect to the rotation axis who measure a net Poynting flux vector.

Here, we study stationary electromagnetic fields on the backgrounds of the
Schwarzschild and Kerr black holes from the point of view of several families of observers.
This approach requires the introduction of sets of reference frames, understood as an idealization
of observers in some state of motion, equipped with some measuring devices. Consequently, the set of reference frames
may be identified with a time-like congruence representing the set of observers' worldlines. In its turn, this congruence is
described by a family of unit time--like vectors, the observers' 4-velocity field, sometimes called the monad field, $\tau$.

As well known, covariant electromagnetic fields can be classified according to their first invariant, $I_1$, as of magnetic, electric, and null types.
The second invariant, $I_2$, generates the pure and non pure classes, see section 2.
Except for pure null fields, it is well known that it is always possible to find the reference frame where the Poynting vector vanishes \cite{Landau}.
Mitskievich showed how to find such a reference frame \cite{Mitskievich06, Classif}. Using the electromagnetic 2-form, he proved that for any non null pure field,  it is always possible to construct a simple bivector from which it is possible to extract a unit time--like vector which represents these
special sets of reference frames.

Whenever the electromagnetic field rotates with the speed of light, a causal barrier appears, the light surface
(or light cylinder, in the case of rigid rotation \cite{Rogava, Osmanov, Osmanov2}). In this case, it becomes
necessary to use two different families of rotating observers, one for each side of the light surface.
The electromagnetic fields on the light surface are always of the pure null type \cite{Nuestroarticulo}.

\medskip

Next, we present the various families of observers in Kerr spacetime that we will consider in this work.
To identify them, as well as the quantities measured in their corresponding reference frame,
we use accents or latin subscripts. These families are

\begin{enumerate}
\item Locally non-rotating observers \cite{Misner}: The world-lines congruence of these observers are not rotating with respect to the local geometry,
        the associated 4-velocity is represented by $\tau$.
        Quantities without accents or subscripts are measured or related to this reference frame.

\item Vanishing Poynting observers: These observers measure no net flux of electromagnetic energy, for them the Poynting vector vanishes.

      \begin{enumerate}
      \item For pure electromagnetic fields($I_2=0$) we consider two subfamilies:
           \begin{enumerate}
                  \item Observers who measure only a magnetic field having a 4-velocity field $\tau_I$.
                  \item Observers who measure only an electric field having a 4-velocity field $\tau_{II}$.
           \end{enumerate}
      \item For non-pure electromagnetic fields($I_2\neq0$) we also consider two subfamilies:
           \begin{enumerate}
                 \item Observers who measure parallel electric and magnetic vectors, obeying $I_1<0$; they  have a 4-velocity field $\tau_{A}$.
                 \item Observers who measure parallel electric and magnetic vectors obeying $I_1>0$;  they have a 4-velocity field $\tau_{B}$.
           \end{enumerate}
      \end{enumerate}

\item Carter observers: In Kerr-Newman spacetime, these observers measure no net flux of electromagnetic energy and gravitational super-energy, for them the Poynting vector and the super-Poynting vector vanishes.
     \begin{enumerate}
            \item World lines of these observers are rotating, their 4-velocity in regions $r<r_-$ and $r_+<r$ is represented by $\tilde{\tau}$, the tilde quantities are measured in this reference frame; for example, $\tilde{\bf E}$ is the electric field measured in this frame and ${\tilde{\theta}^{\alpha}}$ is the tetrad basis associated to this frame.
            \item World lines of this observers are rotating, their 4-velocity in regions $r_-<r<r_+$ is represented by $\check{\tau}$ and the related quantities with check .
     \end{enumerate}

\item  Wylleman observers: These observers also measure no net flux of electromagnetic energy.
          \begin{enumerate}
            \item In Kerr-Newmann spacetime, Wylleman observers, with 4-velocity $\breve{\tau}$, have a relative motion in the radial direction with respect to the rotating Carter observers in regions $r<r_-$ and $r_+<r$, they are mentioned briefly at the end of section 7.
             \item In Kerr-Newmann spacetime, Wylleman observers, with 4-velocity $\bar{\tau}$, have a relative motion in the radial direction with respect to the non-rotating Carter observers in the region $r_-<r<r_+$, they are mentioned briefly at the end of section 7.
              \item General Wylleman observers are briefly discussed in section 8. For electromagnetic fields of the pure electric type, they have the 4-velocity $\tau_E$;
               for fields of the pure magnetic type, they have the 4-velocity $\tau_B$; for fields of the non-pure electric type, they have the four velocity $\tau_e$; and finally, for fields of the non-pure magnetic type, they have the 4-velocity $\tau_b$.
          \end{enumerate}
\end{enumerate}

To consider these observers families we use the theory of arbitrary reference frames, also known as the monad formalism or 3+1 decomposition
\cite{Nova, Antonov, GarciaParrado}.

The paper is organized as follows. Section 2 presents the electromagnetism in arbitrary reference frames, the classification of electromagnetic
fields, and we recall a result on the propagation of electromagnetic fields.
Section 3 introduces the locally non-rotating observers in Kerr spacetime.
Section 4 deals with test stationary axially symmetric electromagnetic fields in Kerr spacetime that are discussed from the point of view of the locally non-rotating observers in Kerr spacetime.
Section 5 presents pure rotating electromagnetic fields described from the point of view of both the locally non-rotating observers and the rotating observers $I$ and $II$, the velocity fields of the last ones are determined with respect to the non-rotating ones.
Section 6 presents non pure electromagnetic fields discussed from the point of view of locally non-rotating observers and the rotating observers $A$ and $B$;
the velocity field of the rotating ones as measured by the non-rotating ones is also presented.
In Section 7, we give some examples of pure test dipolar-like magnetic fields around Schwarzschild and Kerr black holes, provide
graphs of the charge and current densities together with the light surface for each example. Also, the Kerr-Newman electromagnetic field is presented as an example of a non-pure electromagnetic field.
In Section 8, it is shown that vanishing Poynting observers are not unique. As a matter of fact, there is a multitude of them for any
spacetime filled with a non-null electromagnetic field.
Section 9 contains the conclusions.

In this paper, we use the theory of Cartan differential forms \cite{Nuestroarticulo}, the $3+1$--decomposi\-tion or the theory of arbitrary reference frames,
the space-time signature $(+,-,-,-)$ and a system of units in which $c=1$ . Greek indices are taken to run
from 0 to 3 and adopt the standard summation convention on repeated indices.
Furthermore we will represent three vectors with bold symbols, they are 4-dimensional objects in the three dimensional
space orthogonal to the field $\tau$.

\section{Electromagnetism in arbitrary reference fra\-mes}
From the exterior derivative of the electromagnetic covector potential, $A=A_{\mu}(x)dx^\mu$, we get the
electromagnetic field 2-form:
\begin{equation}
F=dA=\frac{1}{2}F_{\mu\nu}dx^\mu\wedge dx^\nu.
\end{equation}
In a reference frame characterized by the monad field $\tau$,
the field tensor decomposes into two terms
\begin{equation}\label{F}
F={\bf E}\wedge \tau + *({\bf B}\wedge \tau)
\end{equation}
where $E$ and $B$ are correspondingly the electric and magnetic covectors measured in that frame \cite{Nova}.

These electric and magnetic fields can also be extracted from the electromagnetic 2-form,
\begin{equation}
{\bf E}= *(\tau\wedge*F)~~~~ {\rm and}~~~~{\bf B}=*(\tau\wedge F)~. \label{EyB}
\end{equation}

In an arbitrary reference frame, the Poynting covector is defined as,
\begin{equation}
{\bf S}=\frac{{\bf E}\times {\bf B}}{4\pi}
= \frac{1}{4\pi}*({\bf E} \wedge \tau \wedge {\bf B})~, \label{Poynting}
\end{equation}
see \cite{Nuestroarticulo}.

The electric current covector, in an arbitrary reference frame, can be
decomposed as:
\begin{equation}
j^\mu={\rho}\tau^\mu+{\bf J^\mu},~~\Rightarrow
~~{\rho}=j\cdot \tau,\,\,{\rm and}\,\,{\bf J^\mu}= b^{\mu\nu}j_\nu~,
\label{j}
\end{equation}
where ${\rho}$ and ${\bf J^\mu}$ are the charge and current densities measured in that arbitrary reference frame.
Here $b_{\mu\nu}$ is the four-dimensional projector, $b=g-\tau\otimes \tau$; it can also be used
as a metric tensor in the physical 3-space orthogonal to $\tau$.
It is also used to define the three-dimensional scalar product $A\bullet B =-b_{\mu\nu}A^\mu B^\nu$,
see \cite{Nuestroarticulo}.
The three-dimensional velocity $\bf v$ of any test particle can be written as:
\begin{equation}
u=\stackrel{\,(\tau)}{u}(\tau+{\bf v})~~~~\Rightarrow~~~~{\bf v}=b\left(\frac{dx}{ds},\cdot\right)~, \label{Fourvelocity}
\end{equation}
where $\stackrel{\,(\tau)}{u}=u\cdot\tau$.

The classification of electromagnetic fields can be established from their invariants (see \cite{Landau} and \cite{Classif}),
\begin{equation}
I_1=-2*(F \wedge *F)=F_{\mu\nu}F^{\mu\nu}=2({\bf B}\bullet {\bf B}-{\bf E}\bullet {\bf E})~,
\end{equation}
\begin{equation}
I_2=2*(F\wedge F)=F^{\,*} _{\mu\nu}F^{\mu\nu}=4{\bf E}\bullet {\bf B}~,
\end{equation}
where
\begin{equation}
F^{~*} _{\mu\nu}=\frac{1}{2}E_{\mu\nu\sigma\tau}F^{\sigma\tau}~.
\end{equation}
According to them, the electromagnetic field can be classified in the following way:
1) Magnetic type if $I_1>0$, 2) Electric type if $I_1<0$), and 3) Null type if  $I_1=0$.
The second invariant introduces the subclassification a) pure if $I_2=0$ and b) not pure if $I_2\neq 0$.

An important result in \cite{Classif} connects the speed of the vanishing Poynting observers with the invariants of the
electromagnetic field:
\begin{equation}
0\leq\frac{|{\bf v}|}{1+{\bf v}^2}=\frac{1}{2}\sqrt{1-\frac{I_1 ^2 + I_2 ^2}{4({\bf E}^2+{\bf B}^2)^2} }=
\frac{|{\bf E}||{\bf B}|}{{\bf E}^2+{\bf B}^2}|\sin\psi|\leq\frac{1}{2}~. \label{PropagationSpeed}
\end{equation}
One can see that for non-null electromagnetic fields, they can always be introduced.

\section{Locally non-rotating observers in Kerr spacetime}
Kerr solution describes a rotating black hole with mass $m$ and angular momentum per unit of rest mass
$a$. In Boyer-Lindquist coordinates \cite{Misner,Stephani,Griffiths}
$$
ds^2= \left(1-\frac{2mr}{\rho^2}\right)dt^2+\frac{4amr\sin^2\vartheta}{\rho^2} dt d\phi
-\frac{\rho^2}{\Delta(r)}dr^2
$$
\begin{equation}
-\rho^2 d\vartheta^2
-\left(r^2+a^2+\frac{2a^2mr\sin^2\vartheta}{\rho^2}\right)\sin^2 \vartheta\,d\phi^2~,
\end{equation}
where
\begin{equation}
\Delta(r) = r^2-2mr+a^2,~~~~\rho^2=r^2+a^2\cos^2\vartheta~.
\end{equation}
From the physical point of view,  acceptable solutions correspond to $m> a^2$; in this case, the Kerr--black hole possesses two horizons corresponding to $\Delta(r)=0$,
\begin{equation}
r_{\pm} = m \pm \sqrt{m^2- a^2}~.
\end{equation}
Kerr black hole also possesses an ergosphere, a region where an observer with 4-velocity $u=u^{0}\partial_t$ cannot be stationary, no matter how powerful their
spaceship is. The four velocity
is a time-like unitary vector satisfying $u\cdot u =g_{00}(u^0)^2=1$. Notice that $g_{00}\leq 0$ in the region $r<r_e$, where
\begin{equation}
r_e=m+\sqrt{m^2-a^2\cos^2\vartheta}.
\end{equation}
This region exists outside the event horizon $r_+$, so a particle can enter it and still can escape without being captured by the black hole.

In what follows we consider only the exterior region outside the exterior horizon $r_+<r$ of the Kerr black hole, or the Schwarzschild
black hole if $a=0$.

We first introduce the following orthonormal tetrad
\begin{equation}
\theta^{(0)}=e^\alpha dt,~~~\theta^{(1)}=e^\beta dr,~~~\theta^{(2)}=e^\gamma d\vartheta~,~~~
\theta^{(3)}=e^\delta  (d\phi-Fdt), \label{KerrNonRotatingBasis}
\end{equation}
where
\begin{equation}
e^{2\alpha}= \frac{\rho^2\Delta}{(r^2+a^2)^2-\Delta a^2\sin^2\vartheta},  ~~~~~e^{2\beta}= \frac{\rho^2}{\Delta},~~~~~
e^{2\gamma}=\rho^2~,
\end{equation}
\begin{equation}
e^{2\delta}=\frac{(r^2+a^2)^2-\Delta a^2\sin^2\vartheta}{\rho^2}\sin^2\vartheta,
~~~~F=a\frac{r^2+a^2-\Delta}{(r^2+a^2)^2-\Delta a^2\sin^2\vartheta}~. \label{metricfunctions}
\end{equation}
This tetrad has a singularity at the horizon $(\Delta = 0)$, but this is only a singularity of the basis  and the coordinate system \cite{Mitskievich2005}.

We consider a family of {\em locally non-rotating observers} around the Kerr black hole; these
observers are not rotating relative to the local spacetime geometry, see exercise 33.3 in \cite{Misner}. The worldlines of these observers are
described by the field
\begin{equation}
\tau=\theta^{(0)}~, \label{MonadaNoRotante}
\end{equation}
which does not rotate since $\omega=*(\theta^{(0)}\wedge d\theta^{(0)})= 0$. Nevertheless,
it is accelerated, $G=-*(\theta^{(0)}\wedge*d\theta^{(0)}) \neq 0$, see \cite{Nuestroarticulo}.

\section{Stationary axially symmetric test-elec\-tro\-mag\-ne\-tic fields in Kerr spacetime}
Let us consider the four potential
\begin{equation}
A=M(r,\vartheta)dt-N(r,\vartheta)d\varphi~. \label{Fourpotential}
\end{equation}
Taking its exterior derivative we find the electromagnetic field:
\begin{equation}
F=\left(M_{,r}dr+M_{,\vartheta}d\vartheta\right)\wedge dt - \left(N_{,r}dr+N_{,\vartheta}d\vartheta\right)\wedge d\varphi~. \label{Fzero}
\end{equation}
In terms of the Kerr non-rotating tetrad (\ref{KerrNonRotatingBasis}), $F$ can be rewritten as follows:
$$
F=\left(M_{,r}-FN_{,r}\right)\hbox{\large e}^{-\alpha-\beta}\theta^{(1)}\wedge\theta^{(0)}
+\left(M_{,\vartheta}-FN_{,\vartheta}\right)\hbox{\large e}^{-\alpha-\gamma}\theta^{(2)}\wedge\theta^{(0)}
$$
\begin{equation}
-N_{,r}\hbox{\large e}^{-\beta-\delta}\theta^{(1)}\wedge\theta^{(3)}
-N_{,\vartheta}\hbox{\large e}^{-\gamma-\delta}\theta^{(2)}\wedge\theta^{(3)}~. \label{F1}
\end{equation}
Other useful forms are
$$
F=*\left[N_{,\vartheta}\hbox{\large e}^{-\gamma-\delta}\theta^{(0)}\wedge\theta^{(1)}
-N_{,r}\hbox{\large e}^{-\beta-\delta}\theta^{(0)}\wedge\theta^{(2)}
\right.
$$
\begin{equation}
\left.
+\left(M_{,\vartheta}-FN_{,\vartheta}\right)\hbox{\large e}^{-\alpha-\gamma}\theta^{(1)}\wedge\theta^{(3)}
-\left(M_{,r}-FN_{,r}\right)\hbox{\large e}^{-\alpha-\beta}\theta^{(2)}\wedge\theta^{(3)}
\right]\label{F2}
\end{equation}
and
$$
F=\left[\left(M_{,r}-FN_{,r}\right)\hbox{\large e}^{-\alpha-\beta}\theta^{(1)}
+\left(M_{,\vartheta}-FN_{,\vartheta}\right)\hbox{\large e}^{-\alpha-\gamma}\theta^{(2)}\right]\wedge\theta^{(0)}
$$
\begin{equation}
-*\left[\left(N_{,\vartheta}\hbox{\large e}^{-\gamma-\delta}\theta^{(1)}
-N_{,r}\hbox{\large e}^{-\beta-\delta}\theta^{(2)}\right)\wedge\theta^{(0)}\right]\label{F3}
\end{equation}
Electric and magnetic covectors, in the locally non-rotating frame, are obtained by comparing (\ref{F3}) with (\ref{F})
\begin{equation}
{\bf E}=\left(M_{,r}-FN_{,r}\right)\hbox{\large e}^{-\alpha-\beta}\theta^{(1)}
+\left(M_{,\vartheta}-FN_{,\vartheta}\right)\hbox{\large e}^{-\alpha-\gamma}\theta^{(2)} \label{Egeneral_non_rotating frame}
\end{equation}
\begin{equation}
{\bf B}=-N_{,\vartheta}\hbox{\large e}^{-\gamma-\delta}\theta^{(1)}
+N_{,r}\hbox{\large e}^{-\beta-\delta}\theta^{(2)}~. \label{Bgeneral_non_rotating frame}
\end{equation}
The Poynting covector (\ref{Poynting}), in the same frame, is
\begin{equation}
{\bf S}=-\frac{1}{4\pi}\left[ N_{,r}\left(M_{,r}-FN_{,r}\right)\hbox{\large e}^{-2\beta}+
N_{,\vartheta}\left(M_{,\vartheta}-FN_{,\vartheta}\right)\hbox{\large e}^{-2\gamma} \right]
\hbox{\large e}^{-\alpha-\delta}\theta^{(3)}~.
\end{equation}
The four current covector, in the non rotating basis, $j=\rho\theta^{(0)}+{\bf J}$, follows from Maxwell equations;
it has the following components:
$$
\rho=-\frac{1}{4\pi}\left\{\left[(M_{,r}-FN_{,r})\hbox{\large e}^{-\alpha-\beta+\gamma+\delta}\right]_{,r}+\right.
$$
\begin{equation}
+\left.
\left[(M_{,\vartheta}-FN_{,\vartheta})\hbox{\large e}^{-\alpha+\beta-\gamma+\delta}\right]_{,\vartheta}
\right\}\hbox{\large e}^{-\beta-\gamma+\delta} \label{chargedensitygeneral}
\end{equation}
$$
{\bf J}=\frac{1}{4\pi}\left[F_{,r}(M_{,r}-FN_{,r})\hbox{\large e}^{-2\alpha-2\beta+\delta}-F_{,\vartheta}(M_{,\vartheta}-FN_{,\vartheta})
\hbox{\large e}^{-2\alpha-2\gamma+\delta}\right.
$$
\begin{equation}
+\left.
\left(N_{,r}\hbox{\large e}^{\alpha-\beta+\gamma-\delta}\right)_{,r}\hbox{\large e}^{-\alpha-\beta-\gamma}
+\left(N_{,\vartheta}\hbox{\large e}^{\alpha+\beta-\gamma-\delta}\right)_{,\vartheta}\hbox{\large e}^{-\alpha-\beta-\gamma}
\right] \theta^{(3)} \label{Currentdensitygeneral}
\end{equation}

The electromagnetic field invariants are
$$
I_1=2\left\{\left(N_{,r}^2\hbox{\large e}^{-2\beta}+N_{,\vartheta}^2\hbox{\large e}^{-2\gamma}\right)\hbox{\large e}^{-2\delta}
-\left[(M_{,r}-FN_{,r})^2\hbox{\large e}^{-2\beta}\right.+
\right.
$$
\begin{equation}
\left.\left.+(M_{,\vartheta}+FN_{\vartheta})^2\hbox{\large e}^{-2\gamma}\right]\hbox{\large e}^{-2\alpha}\right\}
\label{Invariante_1}
\end{equation}
\begin{equation}
I_2=4(M_{,r}N_{,r}-M_{,\vartheta}N_{\vartheta})\hbox{\large e}^{-\alpha-\beta-\gamma-\delta}~.
\end{equation}
Different types of electromagnetic fields may be considered according to the choices of the functions
$M(r,\vartheta)$ and  $N(r,\vartheta)$. Pure electromagnetic rotating fields correspond to either
$M=M(N)$ or $N=N(M)$ and the special case $M=0$ and $a\neq0$ (The Poynting vector arises from the
reference frame dragging). In what follows we consider only the first case $M=M(N)$, since the second one
is quite similar. The cases $M=0$ or $N=0$ are not considered since they result in fields with a
vanishing Poynting vector. Non--pure rotating fields correspond to other choices of $M$ and $N$.

\section{Pure rotating electromagnetic fields in Kerr spacetime}
In this case, the first electromagnetic invariant (\ref{Invariante_1}) has the following expression:
\begin{equation}
I_1=2\left(N_{,r} ^2\hbox{\large e}^{-2\beta-2\delta}+N_{,\vartheta} ^2\hbox{\large e}^{-2\gamma-2\delta}\right)\left[1-\xi(r,\vartheta)^2\right]~,
\label{Inv1puro}
\end{equation}
where
\begin{equation}
\xi(r,\vartheta)=\left(\frac{dM}{dN}-F\right)\hbox{\large e}^{-\alpha+\delta}~. \label{Kerr_xi}
\end{equation}
The electromagnetic field is of pure magnetic type in the region $|\xi|<1$, pure null type on the surface $\xi=1$,
and pure electric type in the region $|\xi|>1$. We shall see below that the rotation speed
of the vanishing Poynting observers is related to the function $\xi$, with
contributions from the electromagnetic field through $dM/dN$ and from the frame dragging in Kerr spacetime through $F$.

In the locally non-rotating frame, the electric and magnetic covectors are obtained from (\ref{Egeneral_non_rotating frame}) and (\ref{Bgeneral_non_rotating frame})
\begin{equation}
{\bf E}=\xi(r,\vartheta)\left(N_{,r}\hbox{\large e}^{-\beta-\delta}\theta^{(1)}
+N_{,\vartheta}\hbox{\large e}^{-\gamma-\delta}\theta^{(2)}\right) \label{Epure_non_rotating frame}
\end{equation}
\begin{equation}
{\bf B}=N_{,\vartheta}\hbox{\large e}^{-\gamma-\delta}\theta^{(1)}
-N_{,r}\hbox{\large e}^{-\beta-\delta}\theta^{(2)} \label{Bpure_non_rotating frame}
\end{equation}

The resulting Poynting covector is
\begin{equation}
{\bf S}=-\frac{1}{4\pi}\xi(r,\vartheta)\left[N_{,r}\hbox{\large e}^{-2\beta}+N_{,\vartheta}\hbox{\large e}^{-2\gamma}\right]\hbox{\large e}^{-2\delta}\theta^{(3)}~.
\end{equation}
The charge density measured by non-rotating observers follows from (\ref{chargedensitygeneral})
\begin{equation}
\rho=-\frac{1}{4\pi}\left\{\left[\xi(r,\vartheta) N_{,r}\hbox{\large e}^{-\beta+\gamma}\right]_{,r}
+
\left[\xi(r,\vartheta)N_{,\vartheta}\hbox{\large e}^{\beta-\gamma}\right]_{,\vartheta}
\right\}\hbox{\large e}^{-\beta-\gamma-\delta}\label{Qdensity_no_rotating}
\end{equation}
while the current density
$$
J=\frac{1}{4\pi}\left\{\left[
\left(N_{,r}\hbox{\large e}^{\alpha-\beta+\gamma-\delta}\right)_{,r}
+\left(N_{,\vartheta}\hbox{\large e}^{\alpha+\beta-\gamma-\delta}\right)_{,\vartheta}
\right]\hbox{\large e}^{-\alpha-\beta-\gamma}+\right.
$$
\begin{equation}
\left.+~\xi(r,\vartheta) \left(F_{,r}N_{,r}\hbox{\large e}^{-\alpha-2\beta}-F_{,\vartheta}N_{,\vartheta}\hbox{\large e}^{-\alpha-2\gamma}\right)\right\}
\theta^{(3)}~,
\label{Jpure_no_rotating}
\end{equation}
is obtained from (\ref{Currentdensitygeneral}).

Since $I_2=0$, the electromagnetic 2-form can be written as a simple bivector. From (\ref{F1}), the electromagnetic field tensor
can be rearranged as
\begin{equation}
F_I=*\left[\left(-N_{,\vartheta}\hbox{\large e}^{-\gamma-\delta}\theta^{(1)}+N_{,r}\hbox{\large e}^{-\beta-\delta}\theta^{(2)}\right)
\wedge\left(\theta^{(0)}-\xi\theta^{(3)}\right)\right],\label{FIa}
\end{equation}
or, from (\ref{F1}), as
\begin{equation}
F_{II}=\xi\left(N_{,r}\hbox{\large e}^{-\beta-\delta}\theta^{(1)}+N_{,\vartheta}\hbox{\large e}^{-\gamma-\delta}\theta^{(2)}\right)
\wedge\left(\theta^{(0)}-\frac{1}{\xi}\theta^{(3)}\right)~.\label{FIIa}
\end{equation}

We introduce in the following two new sets of rotating observers, each one for the magnetic region $|\xi|<1$ and
for the electric region $|\xi|>1$, respectively, which measure no Poynting vector. On the {\it light surface}, $|\xi|=1$, where both electromagnetic invariants vanish, it is not possible to introduce vanishing Poynting observers since they should move with the speed of light, see (\ref{PropagationSpeed}).

\subsection{The pure magnetic type region}
From the field tensor (\ref{FIa}), after the normalization of the time-like covector, we readily find the 4-velocity
of the vanishing Poynting observers:
\begin{equation}
\tau_I=\frac{1}{\sqrt{1-\xi^2}}\left(\theta^{(0)}-\xi\theta^{(3)}\right),~~~|\xi|<1~, \label{tau_kerr_I}
\end{equation}
where the label $I$ is used to distinguish these observers from those mentioned above.

From the point of view of the locally non-rotating observers (\ref{MonadaNoRotante}), these observers are rotating around the Kerr black hole with velocity
\begin{equation}
{\bf v}_I=\xi \theta^{(3)}~. \label{vI}
\end{equation}

Inserting (\ref{FIa})  into
(\ref{EyB}), we find
\begin{equation}
{\bf E}_I=0
\end{equation}
\begin{equation}
{\bf B}_{I}=\sqrt{1-\xi^2}\left(-N_{,\vartheta}\hbox{\large e}^{-\gamma-\delta}\theta^{(1)}+N_{,r}\hbox{\large e}^{-\beta-\delta}\theta^{(2)}\right)~.
\end{equation}
Consequently the Poynting vector vanishes away for observers $I$.

{ In the special case when $u=\tau_I$ is defined everywhere outside the black hole horizon,
the electromagnetic field is of pure magnetic type, $I_1>0$. Moreover, due to the interpretation of $\xi$ as the speed of vanishing Poynting observers $I$ (\ref{vI}), we notice that
(\ref{Epure_non_rotating frame}) may be regarded as a rotation-induced unipolar electric field such that $I_2=4{\bf E\cdot B}=0$.
Hence, the ideal magnetohydrodynamic conditions in \cite{Meier, Sengupta} are satisfied, since the Lorentz force free condition,
\begin{equation}
*(u\wedge*F)=0, \label{LorentzForceFree}
\end{equation}
and Ohm's law,
\begin{equation}
\eta {\bf J}= *(u\wedge*F)=0 \label{ohm},
\end{equation}
are satisfied for a corotating plasma of infinite conductivity. Here $\eta\rightarrow 0$ is the resistivity of the corotating plasma and $u=\tau_I$ is  its four velocity. The fulfillment of (\ref{ohm}) follows from the fact that
${\bf E_I}=0$ and (\ref{EyB}). However, some caution should be taken since these conditions are considered as non causal ones
when time evolution is taken into account \cite{Punsly}.
}

\subsection{The pure electric type region}
From the field tensor (\ref{FIIa}), after the normalization of the time-like covector,  we readily find the 4-velocity of the vanishing
Poynting observers:
\begin{equation}
\tau_{II}=\frac{|\xi|}{\sqrt{\xi^2}-1}\left(\theta^{(0)}-\frac{1}{\xi}\theta^{(3)}\right)~,~~~|\xi|>1~, \label{tau_kerr_II}
\end{equation}
where the label $II$ is used to distinguish these observers from those mentioned above.

From the point of view of the locally non-rotating observers (\ref{MonadaNoRotante}), these observers are rotating around the Kerr black hole with velocity
\begin{equation}
{\bf v}_{II}=\frac{1}{\xi} \theta^{(3)}~. \label{vII}
\end{equation}

Inserting (\ref{FIIa}) into (\ref{EyB}), we find that observers $II$ can only measure an electric field
\begin{equation}
{\bf B}_{II}=0
\end{equation}
\begin{equation}
{\bf E}_{II}=\frac{\sqrt{\xi^2-1}}{|\xi|}\left(N_{,r}\hbox{\large e}^{-\beta-\delta}\theta^{(1)}+N_{,\vartheta}\hbox{\large e}^{-\gamma-\delta}\theta^{(2)}\right)~.
\end{equation}
Consequently, the Poynting vector also vanishes for them.

\section{Non-pure rotating electromagnetic fields}
Now, we have to find the reference frame in which the electric and
magnetic vectors are mutually parallel, which leads to the vanishing of the Poynting vector.
For this purpose, we introduce an auxiliary
2-form \cite{Classif}
\begin{equation}
{\cal F} = \cos(\alpha) F + \sin(\alpha) *F~, \label{dual}
\end{equation}
where $\alpha(x)$ is an arbitrary function chosen in such a way that $\cal F$ becomes
either a simple bivector or the dual conjugation of a simple bivector.

The original electromagnetic field tensor can be readily obtained from
\begin{equation}\label{impuro2}
F = \cos(\alpha) {\cal F} - \sin(\alpha) *{\cal F}~.
\end{equation}
Using the following auxiliary covectors
\begin{equation}
{\cal E}={\cal E}_{(\mu)}\theta^{(\mu)}=*(\theta^{(0)}\wedge*{\cal F})~~~{\rm  and }~~~
{\cal B}={\cal B}_{(\mu)}\theta^{(\mu)}=*(\theta^{(0)}\wedge {\cal F})~,\label{EyBaux}
\end{equation}
we can rewrite ${\cal F}$ as
\begin{equation}
{\cal F}={\cal E}\wedge\theta^{(0)}+*\left({\cal B}\wedge\theta^{(0)}\right)~. \label{FormaAuxiliar}
\end{equation}
Furthermore, we introduce the following two invariants
\begin{equation}
{\cal I}_1=-2*({\cal F}\wedge *{\cal F})=2\left({\cal B}\cdot{\cal B}-{\cal E}\cdot{\cal E}\right)~, \label{InvarAux1}
\end{equation}
\begin{equation}
{\cal I}_2=2*({\cal F}\wedge {\cal F})=4{\cal E}\cdot{\cal B}~, \label{InvarAux2}
\end{equation}
where the latter is in fact a pseudo invariant, since it changes sign under coordinate transformations.

The auxiliary 2-form (\ref{FormaAuxiliar}) can be rewritten as \cite{Kosyakov,Arrayas}
\begin{equation}
{\cal F}=\frac{1}{{\cal E}_{(1)}}{\cal E}\wedge
\left( {\cal E}_{(1)}\theta^{(0)}+ {\cal B}_{(3)}\theta^{(2)}-{\cal B}_{(2)}\theta^{(3)} \right)
-\frac{\cal E \cdot B}{{\cal E}_{(1)}}\theta^{(2)}\wedge\theta^{(3)}~,\label{Aux1}
\end{equation}
\begin{equation}
{\cal F}=\frac{1}{{\cal E}_{(2)}}{\cal E}\wedge
\left( {\cal E}_{(2)}\theta^{(0)}+ {\cal B}_{(1)}\theta^{(3)}-{\cal B}_{(3)}\theta^{(1)} \right)
-\frac{\cal E \cdot B}{{\cal E}_{(2)}}\theta^{(3)}\wedge\theta^{(1)}~,\label{Aux2}
\end{equation}
\begin{equation}
{\cal F}=\frac{1}{{\cal E}_{(3)}}{\cal E}\wedge
\left( {\cal E}_{(3)}\theta^{(0)}- {\cal B}_{(1)}\theta^{(2)}+{\cal B}_{(2)}\theta^{(1)} \right)
-\frac{\cal E \cdot B}{{\cal E}_{(3)}}\theta^{(1)}\wedge\theta^{(2)}~,\label{Aux3}
\end{equation}

The following simple, but important, lemma can be stated.\\
\noindent
{\bf Lemma.} The 2-form $\cal F$ is a simple bivector if and only if the second invariant vanishes
${\cal I}_2=0$.

{\bf Proof:}
\begin{itemize}

\item If ${\cal F}= p\wedge q$ then ${\cal I}_2=*(p\wedge q\wedge p \wedge q)\equiv 0$.
\item If ${\cal I}_2={\cal E}\cdot{\cal B}=0$ then ${\cal F}=p\wedge q$, using (\ref{Aux1})-(\ref{Aux3}).
\end{itemize}
$\square$

Consequently, putting ${\cal I}_2=0$ we can express ${\cal F}$ as a simple bivector using either (\ref{Aux1}) or (\ref{Aux2}),
\begin{equation}
{\cal F}={\cal E}\wedge
\left( \theta^{(0)}-\frac{{\cal B}_{(2)}}{{\cal E}_{(1)}}\theta^{(3)} \right)\label{impuro3}
={\cal E}\wedge
\left(\theta^{(0)}+ \frac{{\cal B}_{(1)}}{{\cal E}_{(2)}}\theta^{(3)} \right)~.
\end{equation}
Since
\begin{equation}
{\cal I}_2= 4{\cal E}\cdot{\cal B}=4\left({\cal E}_{(1)}{\cal B}_{(1)}+{\cal E}_{(2)}{\cal B}_{(2)}\right)=0
~~~\Rightarrow~~~\frac{{\cal B}_{(1)}}{{\cal E}_{(2)}}=
-\frac{{\cal B}_{(2)}}{{\cal E}_{(1)}} \label{I2}~,
\end{equation}
due to the fact that ${\cal E}_{(3)}={\cal B}_{(3)}=0$ , see (\ref{CalE}) and (\ref{CalB}).

The auxiliary covectors can be readily found inserting (\ref{dual}) in (\ref{EyBaux}); they become
\begin{equation}
{\cal E}= \cos\alpha E -\sin \alpha B~,\label{CalE}
\end{equation}
\begin{equation}
{\cal B}= \sin\alpha E +\cos \alpha B~.\label{CalB}
\end{equation}
Here $E$ and $B$  are given by (\ref{Egeneral_non_rotating frame}) and (\ref{Bgeneral_non_rotating frame}).
They are the corresponding non-pure electric and magnetic field vectors measured in the
locally non-rotating reference frame.

Substituting (\ref{Egeneral_non_rotating frame}) and (\ref{Bgeneral_non_rotating frame}) into (\ref{CalE}) and (\ref{CalB}), the non-null components of the auxiliary covectors in the tetrad (\ref{KerrNonRotatingBasis}) are
\begin{equation}
{\cal E}_{(1)}=\cos(\alpha)(M_{,r}-FN_{,r}) \hbox{\large e}^{-\alpha-\beta}+\sin(\alpha)N_{,\vartheta}\hbox{\large e}^{-\gamma-\delta}~,
\end{equation}
\begin{equation}
{\cal E}_{(2)}=\cos(\alpha)(M_{,\vartheta}-FN_{,\vartheta}) \hbox{\large e}^{-\alpha-\gamma}-\sin(\alpha)N_{,r}\hbox{\large e}^{-\beta-\delta}~,
\end{equation}
\begin{equation}
{\cal B}_{(1)}=\sin(\alpha)(M_{,r}-FN_{,r}) \hbox{\large e}^{-\alpha-\beta}-\cos(\alpha)N_{,\vartheta}\hbox{\large e}^{-\gamma-\delta}~,
\end{equation}
\begin{equation}
{\cal B}_{(2)}=\sin(\alpha)(M_{,\vartheta}-FN_{,\vartheta}) \hbox{\large e}^{-\alpha-\gamma}+\cos(\alpha)N_{,r}\hbox{\large e}^{-\beta-\delta}~.
\end{equation}
Inserting (\ref{dual}) in (\ref{InvarAux1}) and (\ref{InvarAux2}), the auxiliary invariants
can be expressed in terms of the electromagnetic ones
\begin{equation}
{\cal I}_1=\cos(2\alpha)I_1+\sin(2\alpha)I_2,~~~{\cal I}_2=\cos(2\alpha) I_2 -\sin(2\alpha) I_1~.
\end{equation}
Hence, putting ${\cal I}_2=0$, the function $\alpha$ becomes
\begin{equation}\label{alpha}
\tan(2\alpha)=I_2/I_I~.
\end{equation}
Notice that $-\pi/4<\alpha<\pi/4$ for $I_1\neq0$.
The first auxiliary invariant becomes
\begin{equation}
{\cal I}_1=\frac{1}{\cos(2\alpha)}I_1~.
\end{equation}
It has the same sign as $I_1$, since $\cos(2\alpha)>0$. Consequently, non-pure electromagnetic fields of the electric type correspond to  ${\cal I}_1<0$,
while those of the magnetic type correspond to  ${\cal I}_1>0$.

\subsection{Non-pure fields of the electric type}
Non-pure fields of the electric type can be obtained from (\ref{impuro3}), since its substitution in (\ref{InvarAux1}) results in ${\cal I}_1<0$.

Thus, for the 2-form of field (\ref{impuro3}), in the case when $|{\cal E}_{(1)}|>|{\cal B}_{(2)}|$, we can choose the monad field as
\begin{equation}
\tau_{A}=\frac{|{\cal E}_{(1)}|}{ \sqrt{ {\cal E}_{(1)} ^2 - {\cal B}_{(2)} ^2 } }
\left( \theta^{(0)}-\frac{{\cal B}_{(2)}}{{\cal E}_{(1)}}\theta^{(3)} \right)~,\label{Monad1Impure}
\end{equation}
where the subscript $A$ is used to distinguish the observers related to this monad field from the ones already
mentioned above.

Comparing (\ref{Monad1Impure}) with (\ref{Fourvelocity}), one finds that in the locally non-rotating reference frame (\ref{MonadaNoRotante}) the velocity field of the vanishing Poynting observers is given by
\begin{equation}
v_{A}=\frac{{\cal B}_{(2)}}{{\cal E}_{(1)}}. \label{V1impure}
\end{equation}
Inserting (\ref{impuro3}) into (\ref{impuro2}), we have
\begin{equation}
F=
\left[ \cos(\alpha) \left(p\wedge \tau_{A}\right)
- \sin(\alpha) *\left(p\wedge
\tau_{A}\right)\right]~,
\end{equation}
where
\begin{equation}
p=\frac{\sqrt{ {\cal E}_{(1)} ^2 - {\cal B}_{(2)} ^2 } }{|{\cal E}_{(1)}|}~
{\cal E}~.
\end{equation}
The expressions for the electric and magnetic fields in the frame (\ref{Monad1Impure}) are readily obtained by comparing
(\ref{impuro3}) with (\ref{F}),
\begin{equation}
{\bf E}_{A}= \cos(\alpha) p, ~~~~{\bf B}_{A}=-\sin(\alpha) p~,
\end{equation}
both covectors are parallel to each other, consequently the Poynting vector vanishes in this reference frame.

\subsection{Non-pure fields of the magnetic type}
In the opposite case, $|{\cal B}_{(2)}|>|{\cal E}_{(1)}|$, the auxiliary 2-form can be rewritten as
\begin{equation}
{\cal F}=*
\left[{\cal B}
\wedge \left( \theta^{(0)}-\frac{{\cal E}_{(1)}}{{\cal B}_{(2)}}\theta^{(3)} \right)
\right]~. \label{impuro4}
\end{equation}
Its substitution in (\ref{InvarAux1}) results in ${\cal I}_1>0$, which corresponds to the non-pure magnetic type.

On the other hand, inserting (\ref{impuro4}) into (\ref{impuro2}), leads to
\begin{equation}
F = \sin(\alpha)\left(q\wedge\tau_{B} \right) +\cos(\alpha)*\left(q\wedge\tau_{B} \right)~, \label{impuro5}
\end{equation}
where
\begin{equation}
q=
\frac{ \sqrt{ {\cal B}_{(2)} ^2 - {\cal E}_{(1)} ^2 } }{|{\cal B}_{(2)}|}~{\cal B}
\end{equation}
and
\begin{equation}
\tau_{B} = \frac{|{\cal B}_{(2)}|}{ \sqrt{ {\cal B}_{(2)} ^2 - {\cal E}_{(1)} ^2 } }
\left( \theta^{(0)}-\frac{{\cal E}_{(1)}}{{\cal B}_{(2)}}\theta^{(3)} \right)~,
\label{Monad2Impure}
\end{equation}
here the subscript $B$ is used to distinguish the observers related to this monad field from the ones already
mentioned above.

Comparing (\ref{Monad2Impure}) with (\ref{Fourvelocity}), one finds that the velocity field of the vanishing Poynting observers
in the locally non-rotating reference frame (\ref{MonadaNoRotante}) is given by
\begin{equation}
v_{B}=\frac{{\cal E}_{(1)}}{{\cal B}_{(2)}}. \label{V2impure}
\end{equation}

The expressions for the electric and magnetic fields in the frame (\ref{Monad2Impure}) are readily obtained by comparing
(\ref{impuro3}) with (\ref{F}),
\begin{equation}
E_{B}= \sin(\alpha) q~, ~~~~B_{B}=\cos(\alpha) q~.
\end{equation}
Since both covectors are parallel to each other, one has a zero Poynting vector in this frame.

\section{Examples of electromagnetic fields around black holes}

\subsection{Pure fields}
In this section, we present some graphics of charge and electric current densities
associated to rotating dipolar-like test magnetic fields around
black holes. In order to see the nature of the electromagnetic field around
the black hole we also present graphics of the function
$\xi$, according to which the electromagnetic field is of the pure magnetic type if $\xi<1$,
of the pure null type if $\xi=1$ and of the pure
electric type if $\xi>1$, see equation~(\ref{Inv1puro}).
Since these configurations are stationary ones, they may model the
final evolution stages of accretion disks around black holes,
once the infalling matter supply has been depleted.

In order to consider dipolar like fields immediately outside
the black hole's horizon, $r>r_+$, let us assume for $N(r,\vartheta)$ the following form
\begin{equation}
N(r,\vartheta)=\frac{k \Delta^{5/2} \sin^2\vartheta}{r^6}~, \label{N}
\end{equation}
where $k$ is a constant that could be interpreted as the dipolar moment.

Inserting (\ref{N}) into (\ref{Bpure_non_rotating frame}), the magnetic
field in the locally non-rotating reference frame (\ref{MonadaNoRotante}) is
$$
{\bf B}=-\frac{\Delta^{5/2}}{r^5\sqrt{(r^2+a^2)^2-a^2\Delta\sin^2\vartheta}}
\left\{\left[\frac{2k\cos\vartheta}{r}\theta^{(1)}
+\frac{k\sqrt{\Delta}\sin\vartheta}{ r^2}
\theta^{(2)}\right]\right.
$$
\begin{equation}
\left.-\frac{5(mr-a^2)\sin\vartheta}{\sqrt{\Delta} r^2}\theta^{(2)}\right\}~.
\end{equation}
This field vanishes on the event horizon $r=r_+$, since $\Delta=0$, and behaves as that of a point magnetic
dipole far away from it:
\begin{equation}
{\bf B} \cong -\left[ \frac{2k \cos\vartheta}{r^3}\theta^{(1)} + \frac{k \sin\vartheta}{r^3}\theta^{(2)} \right] ,
\end{equation}
hence $k$ can be interpreted as the dipolar moment.
Inserting (\ref{N}) into (\ref{Epure_non_rotating frame}), the unipolar induced electric field in the locally non-rotating field (\ref{MonadaNoRotante})
is
$$
{\bf{E}}=\frac{\xi\,\Delta^{5/2}}{r^5\sqrt{(r^2+a^2)^2-a^2\Delta\sin^2\vartheta}}
\left\{\left[
-\frac{k\sqrt{\Delta}\sin\vartheta}{r^2}\theta^{(1)}+\frac{2k\cos\vartheta}{r}\theta^{(2)}\right]
\right.
$$
\begin{equation}
+\left.\frac{5(mr-a^2)\sin\vartheta}{\sqrt{\Delta}r^2}\theta^{(1)}\right\}~.
\end{equation}

Function $M(N)$ (or its derivative) should be determined from
observations of the structure of the black hole's
magnetosphere. However, since there is little reliable information,
we present two examples, which differ by the choice of the derivative of the function $M$.

\bigskip

\noindent
{\bf Example 1:}
Let us consider the function (\ref{N}) with the constant derivative
\begin{equation}
\frac{dM}{dN}=\Omega_0~,\label{derivative1}
\end{equation}
For plotting, we use $\Omega_0=1$, $k=1$, $m=1$, and $a\in\{0, 0.5, 0.9\}$.
The profiles of revolution of the current density, in the locally non-rotating reference frame, (\ref{Jpure_no_rotating}), are presented in Figure 1,
while those of the charge density, (\ref{Qdensity_no_rotating}) are shown in Figure 3. The nature of the electromagnetic field (its type)
is shown using the profiles of revolution of the function $\xi$: pure magnetic field corresponds to lines where $\xi<0$, pure null field to lines where
$\xi=0$ and pure electric field to those lines where $\xi>1$, see (\ref{Inv1puro}).
The light surface $\xi=1$, with electromagnetic fields rotating at the speed of light, is displayed in Figure 5.

\bigskip

\noindent
{\bf Example 2:} Let us consider the function (\ref{N}) with exponential derivative
\begin{equation}
\frac{dM}{dN}=\Omega_0\exp\left(-\sigma/N^2\right)~.
\end{equation}
The numerical values that we use in this case are: $\Omega_0=1$, $k=20$, $\sigma=0.1$, $m=1$, and $a\in\{0, 0.5, 0.9\}$.
The same quantities as in the previous case are presented in Figures 2, 4, and 6, respectively.

\begin{figure*}
  \begin{tabular}{ccccc}
   $a=0$ & $a=0.5$  & $a=0.9$ \\
   \includegraphics[width=0.33\textwidth]{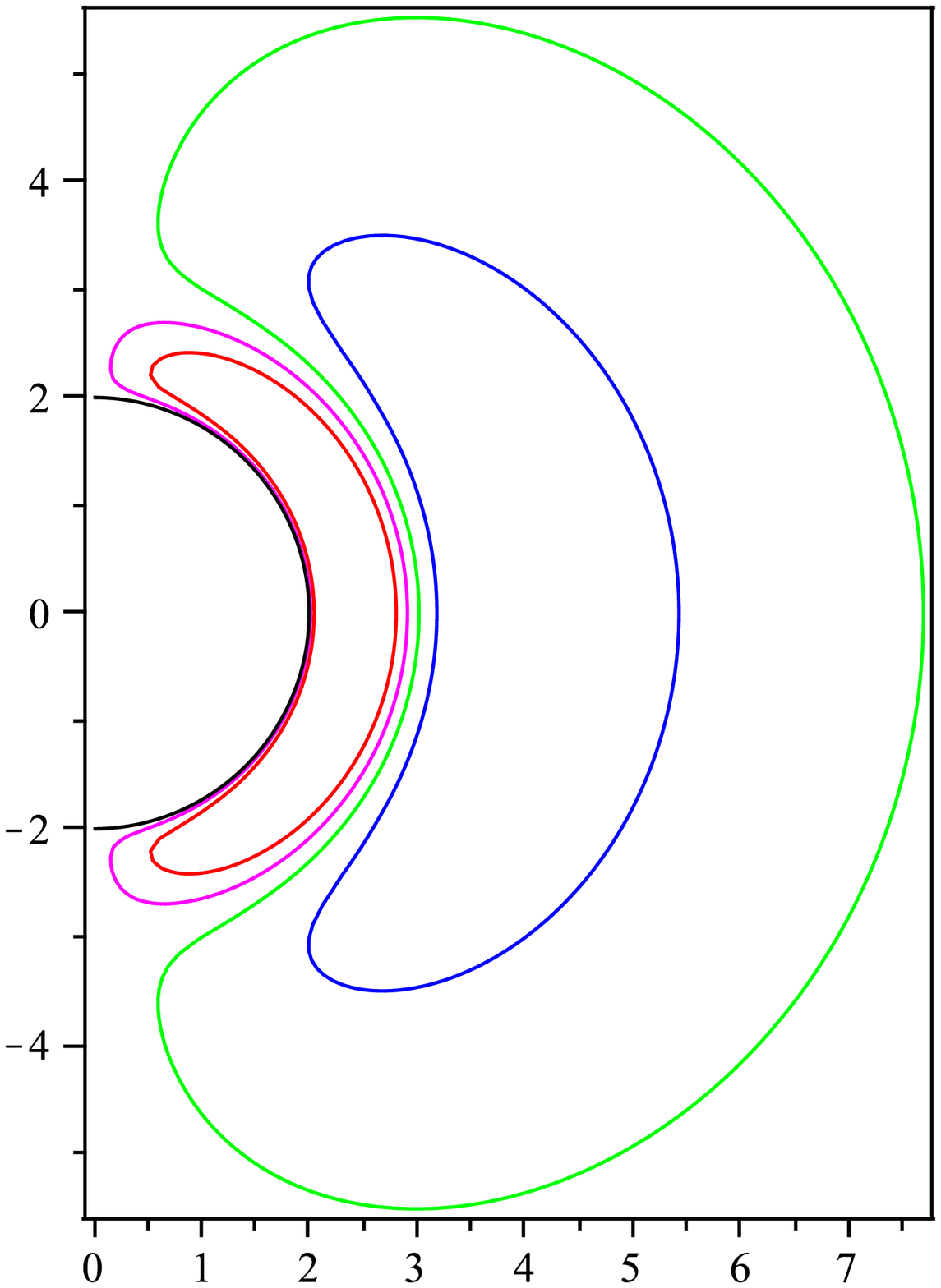} &
   \includegraphics[width=0.33\textwidth]{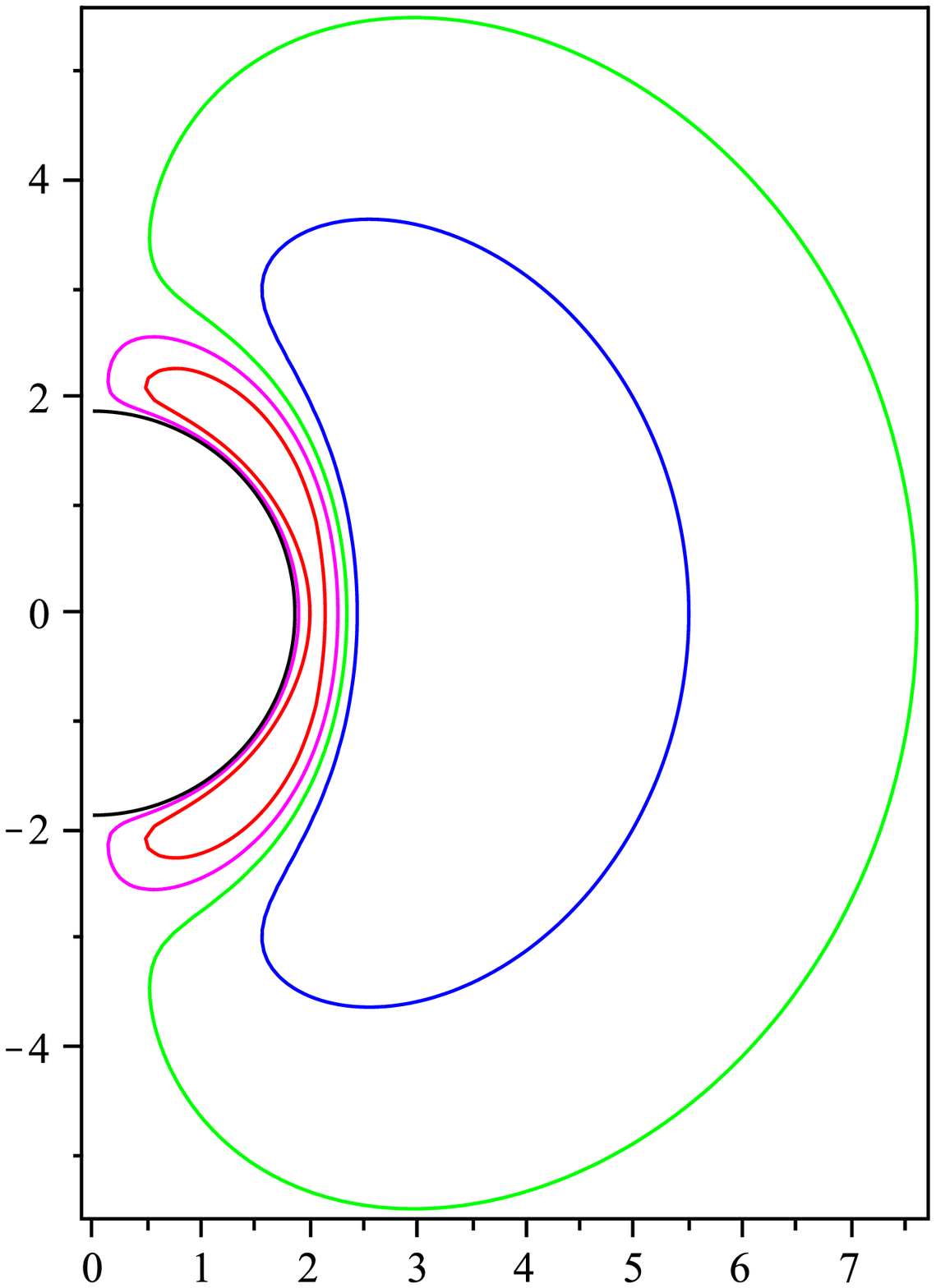} &
   \includegraphics[width=0.33\textwidth]{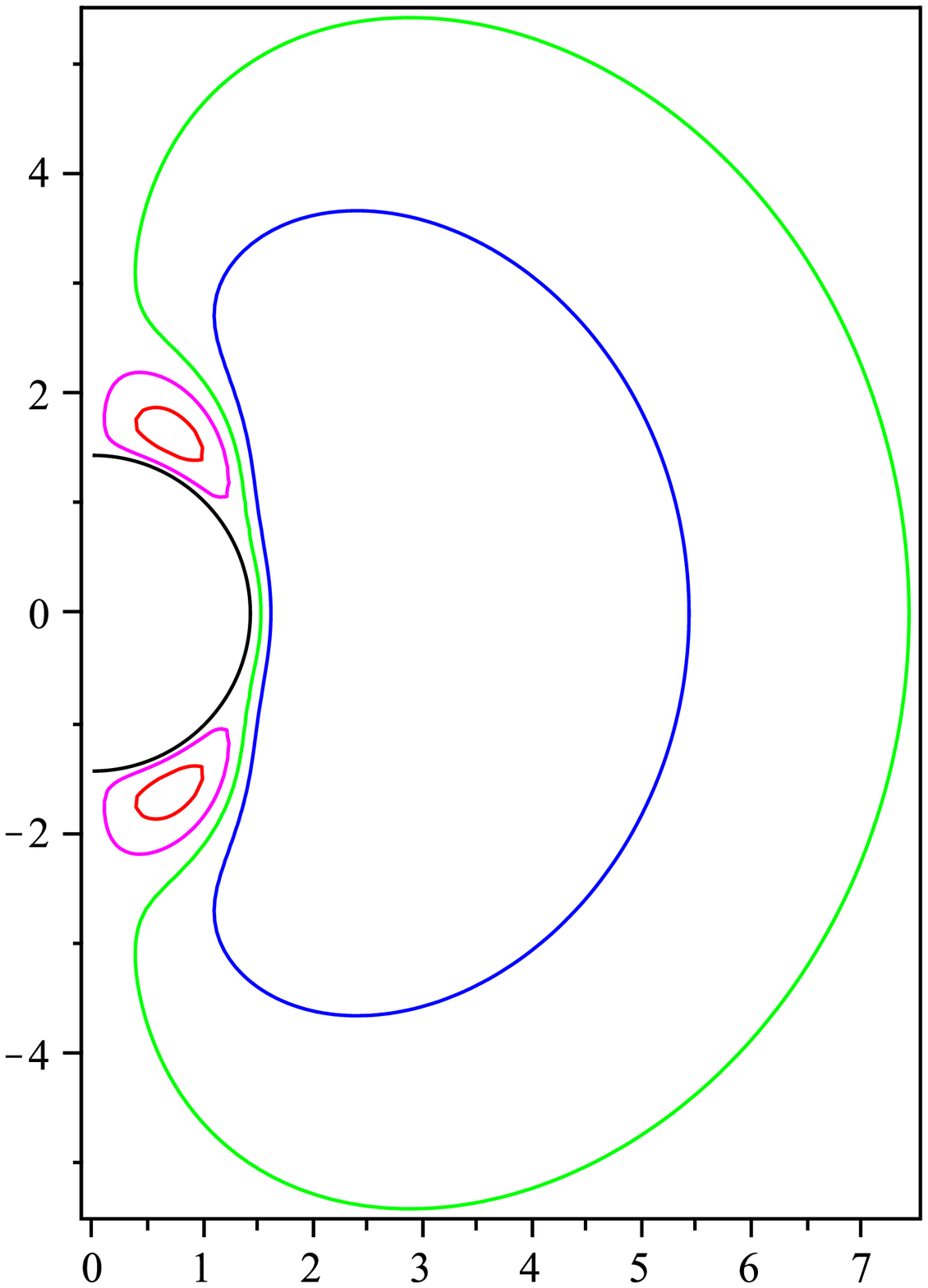}
 \end{tabular}
 \caption{{\footnotesize Contours of profiles of revolution around the vertical axis (rotation axis) depicting constant current density lines (\ref{Jpure_no_rotating}) corresponding to Example 1 for three different values of the Kerr's parameter $a$.
 The red curve corresponds to
 $J=0.0001$, blue to $J=-0.0001$, magenta to $J=0.00003$, and
 green to $J=-0.00003$. The black circle is the black hole's horizon.}}
\end{figure*}

\begin{figure*}
\centering
  \begin{tabular}{ccccc}
   $a=0$ & $a=0.5$  & $a=0.9$ \\
   \includegraphics[width=0.33\textwidth]{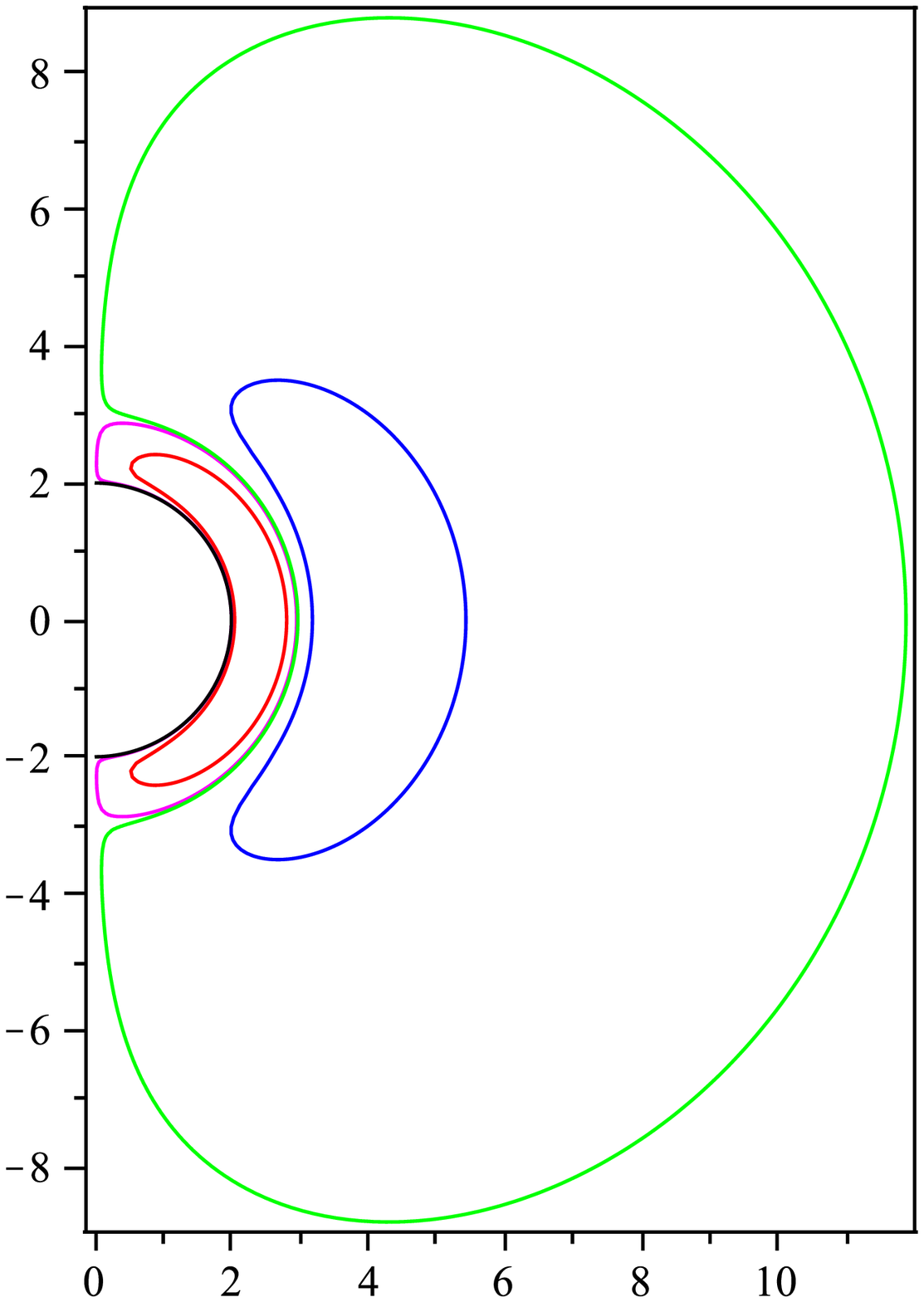} &
   \includegraphics[width=0.33\textwidth]{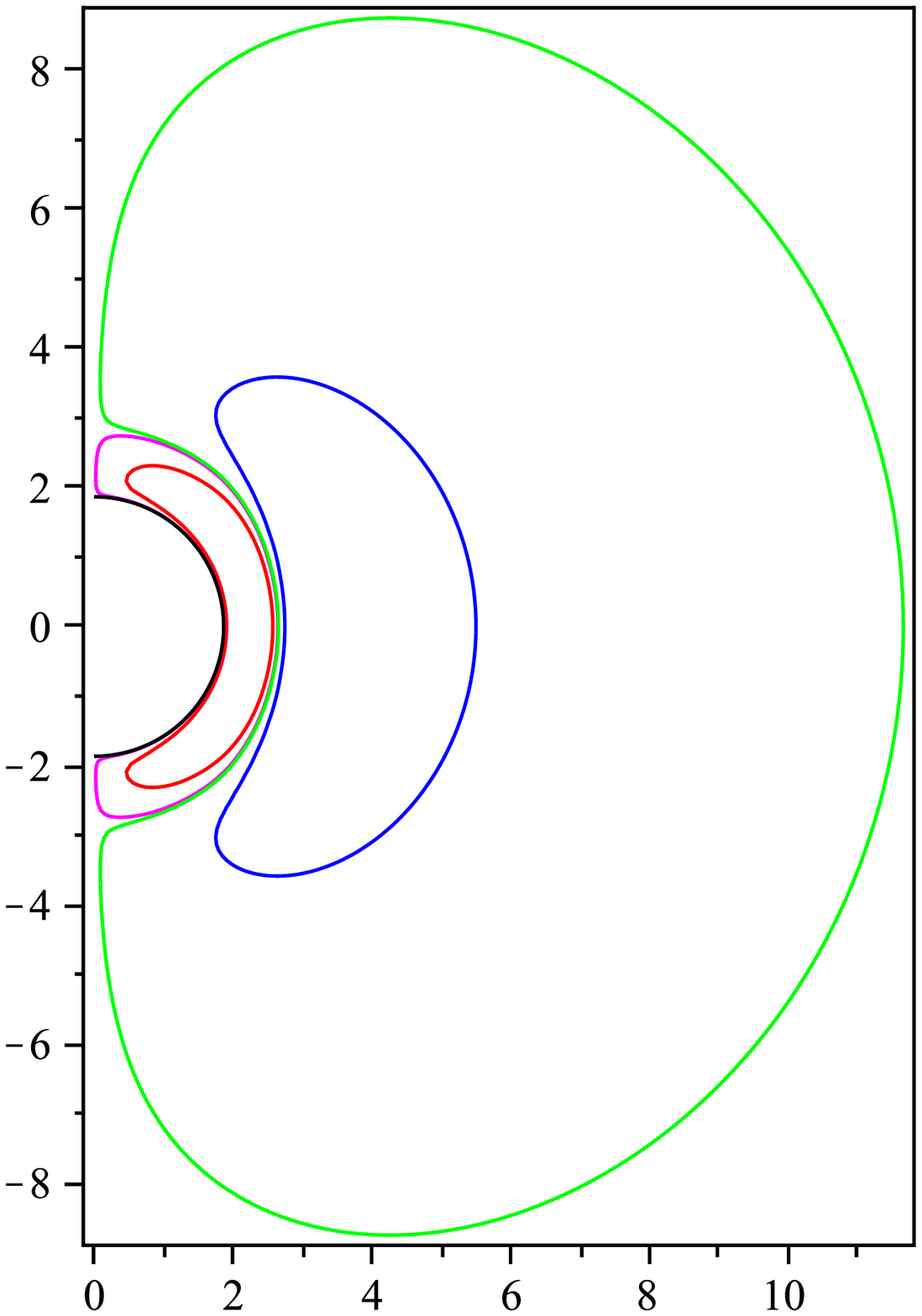} &
   \includegraphics[width=0.33\textwidth]{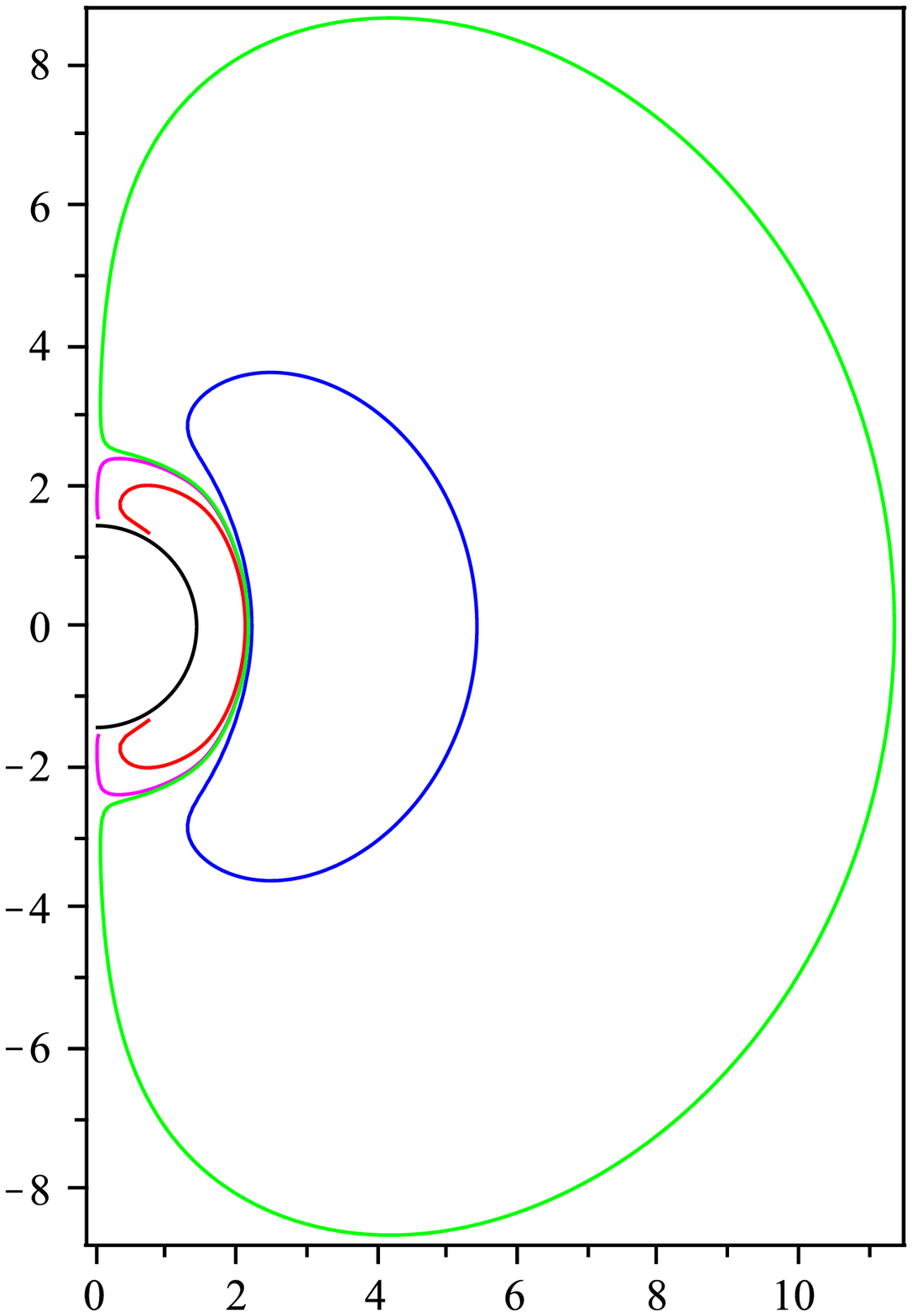}
 \end{tabular}
 \bigskip
 \caption{{\footnotesize
 Same type of contours as in the previous figure, but corresponding to Example 2 for three different values of the Kerr's parameter $a$. The same sequence of colors correspond to $J=0.002$, $J=-0.002$, $J=0.0001$, and $J=-0.0001$, respectively, while the black hole's horizon is the black circle.}}
\end{figure*}

\begin{figure*}
 \centering
 \begin{tabular}{ccccc}
   $a=0$ & $a=0.5$ & $a=0.9$ \\
   \includegraphics[width=0.35\textwidth]{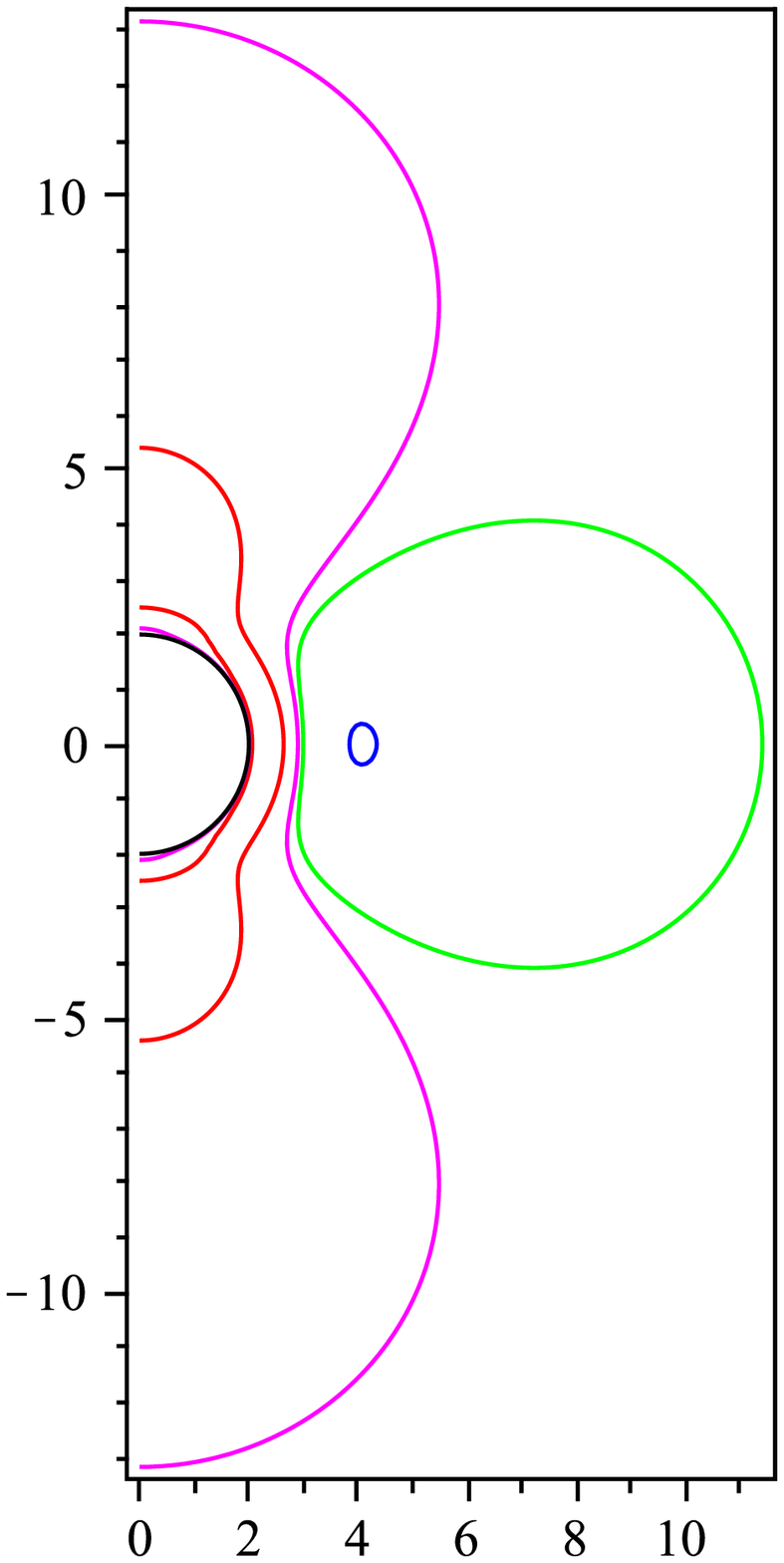} &
   \includegraphics[width=0.35\textwidth]{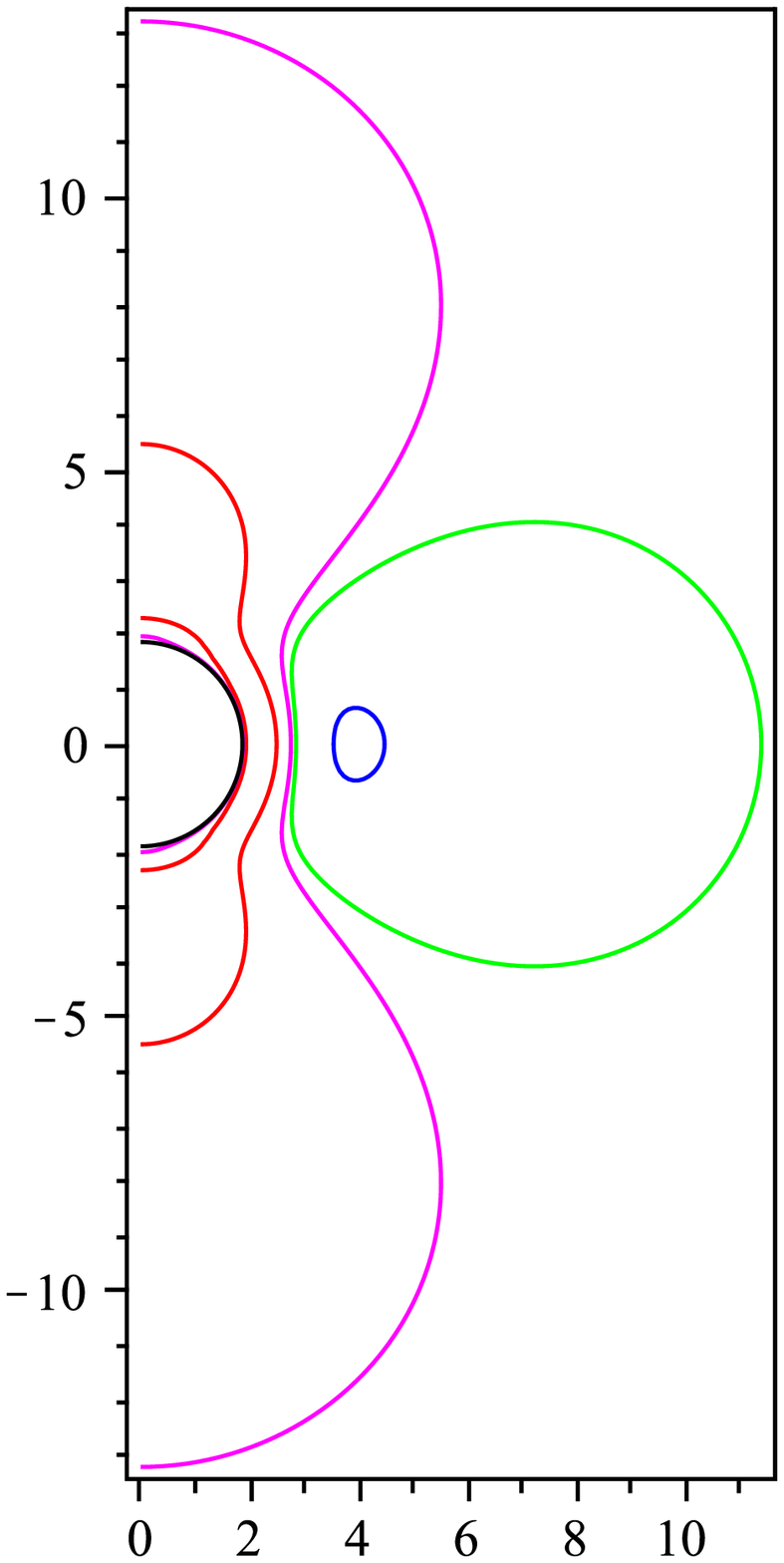} &
   \includegraphics[width=0.35\textwidth]{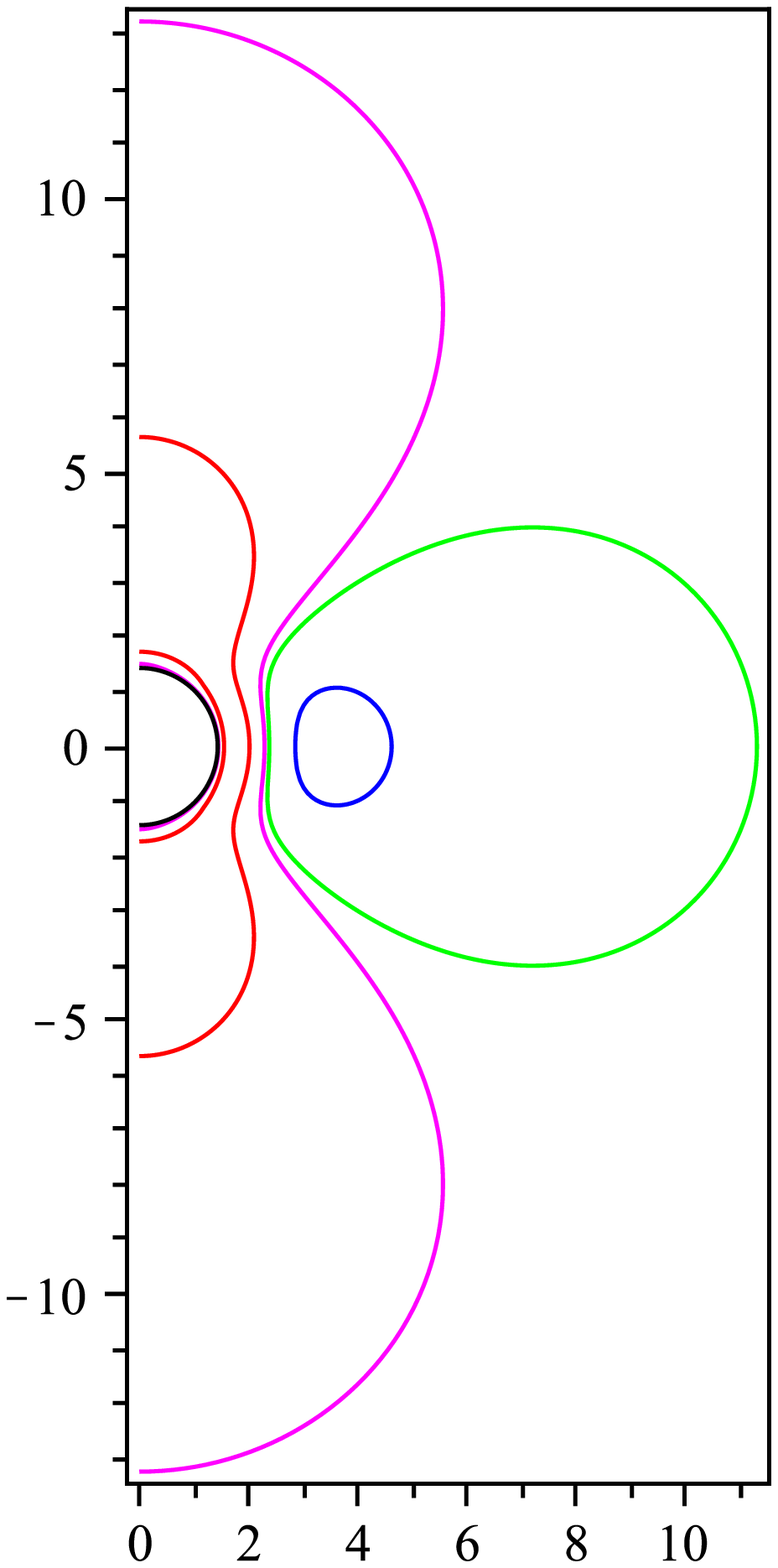}
 \end{tabular}
 \bigskip
 \caption{{\footnotesize
 Contour plots of constant charge density lines according to (\ref{Qdensity_no_rotating}).
 They correspond to the assumptions given in Example 1 for three different values of the Kerr's parameter $a$.
 Red lines are for $\rho=0.0008$, blue for $\rho=-0.0008$, magenta for $\rho=0.0001$, and green for$\rho=-0.0001$; the black curve
 represents the black hole's horizon. } }
\end{figure*}

\begin{figure*}
 \centering
 \begin{tabular}{ccccc}
   $a=0$ & $a=0.5$ & $a=0.9$ \\
   \includegraphics[width=0.35\textwidth]{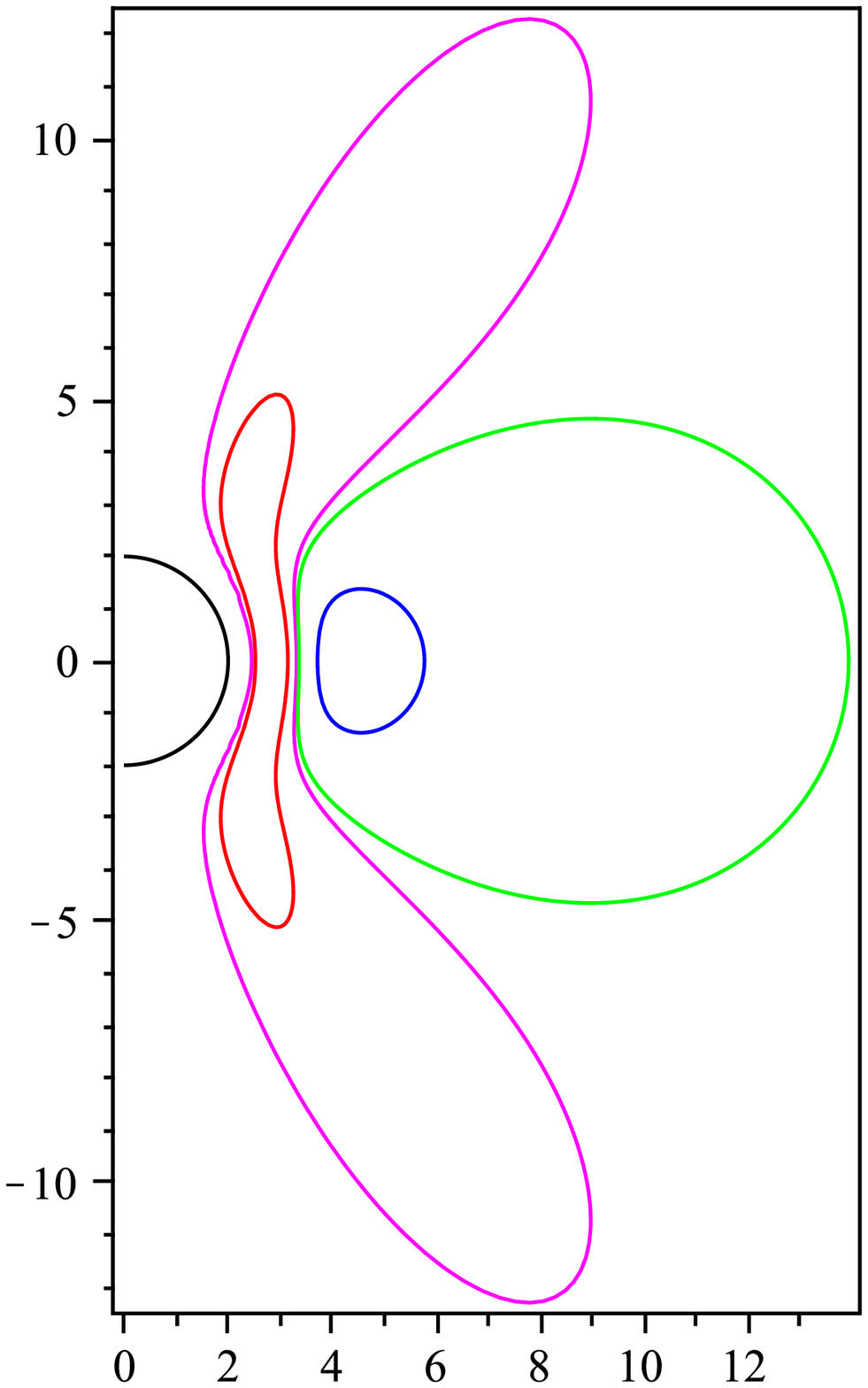} &
   \includegraphics[width=0.35\textwidth]{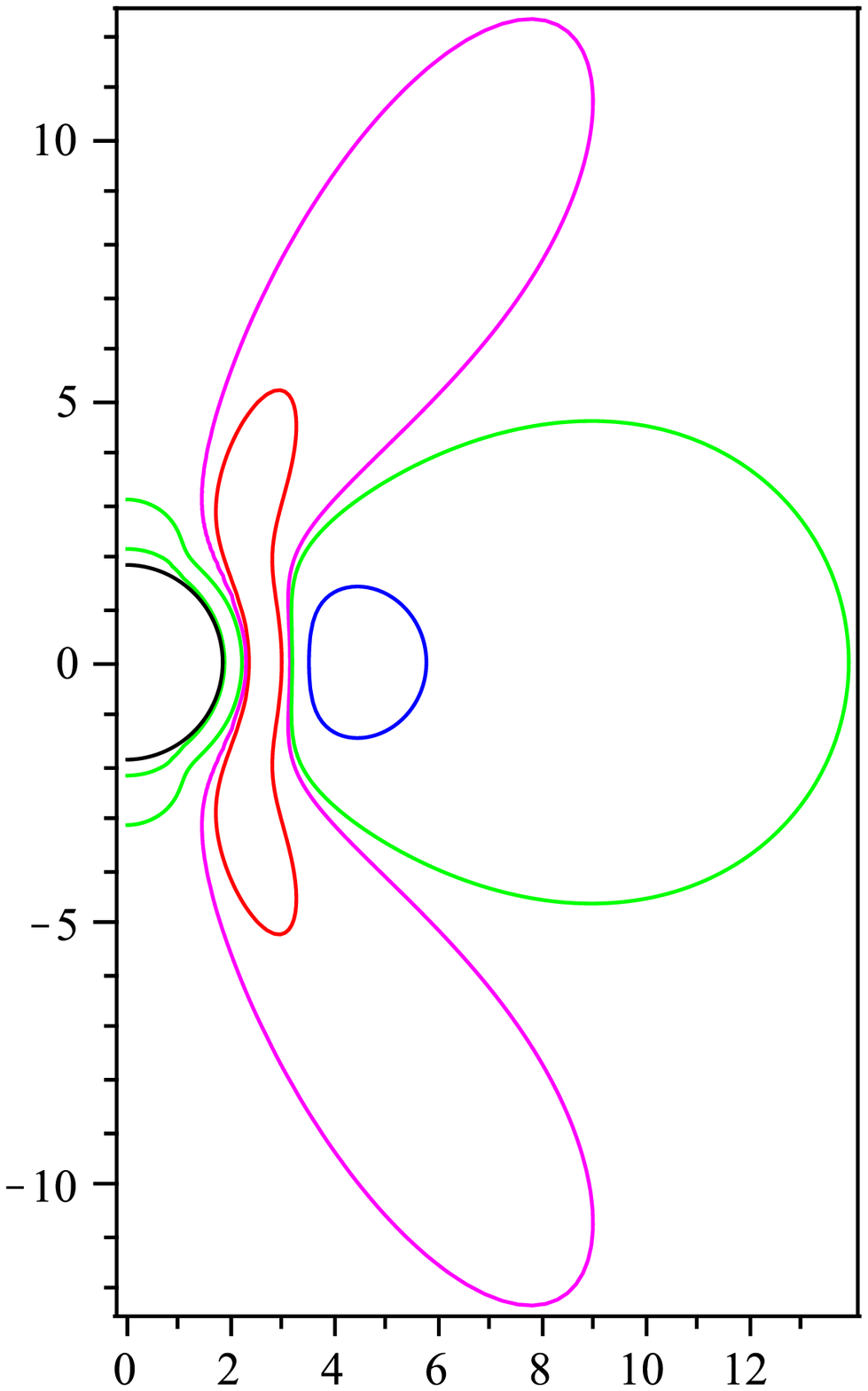} &
   \includegraphics[width=0.35\textwidth]{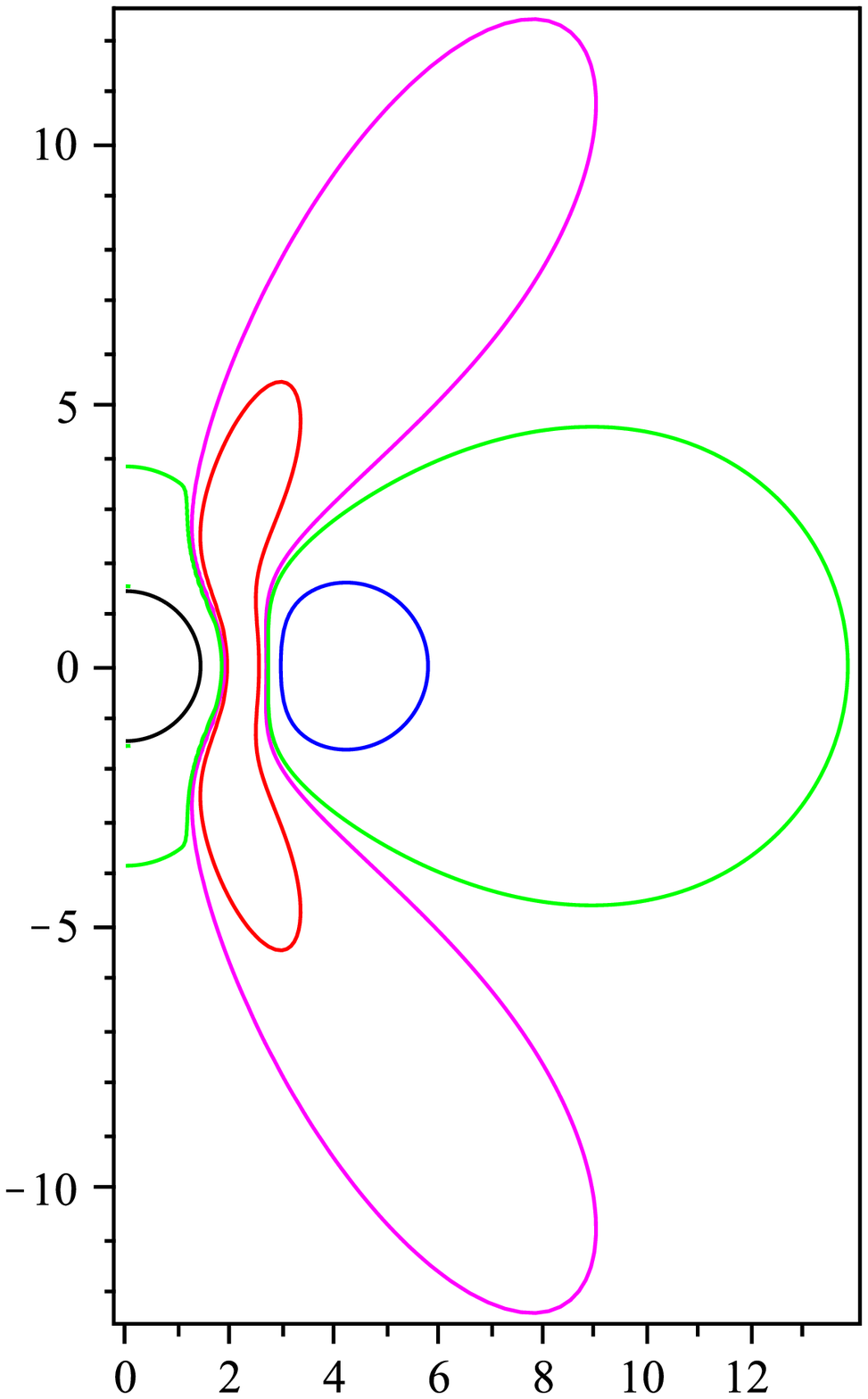}
 \end{tabular}
 \bigskip
 \caption{{\footnotesize
 The same as in the previous figure, but
  corresponding to Example 2.
  Red lines are for $\rho=0.01$, blue for $\rho=-0.01$, magenta for $\rho=0.001$, and green for$\rho=-0.001$; the black curve
 stands for the black hole's horizon. }}
\end{figure*}

\begin{figure*}
 \centering
 \begin{tabular}{ccccc}
   $a=0$ & $a=0.5$  & $a=0.9$ \\
   \includegraphics[width=0.35\textwidth]{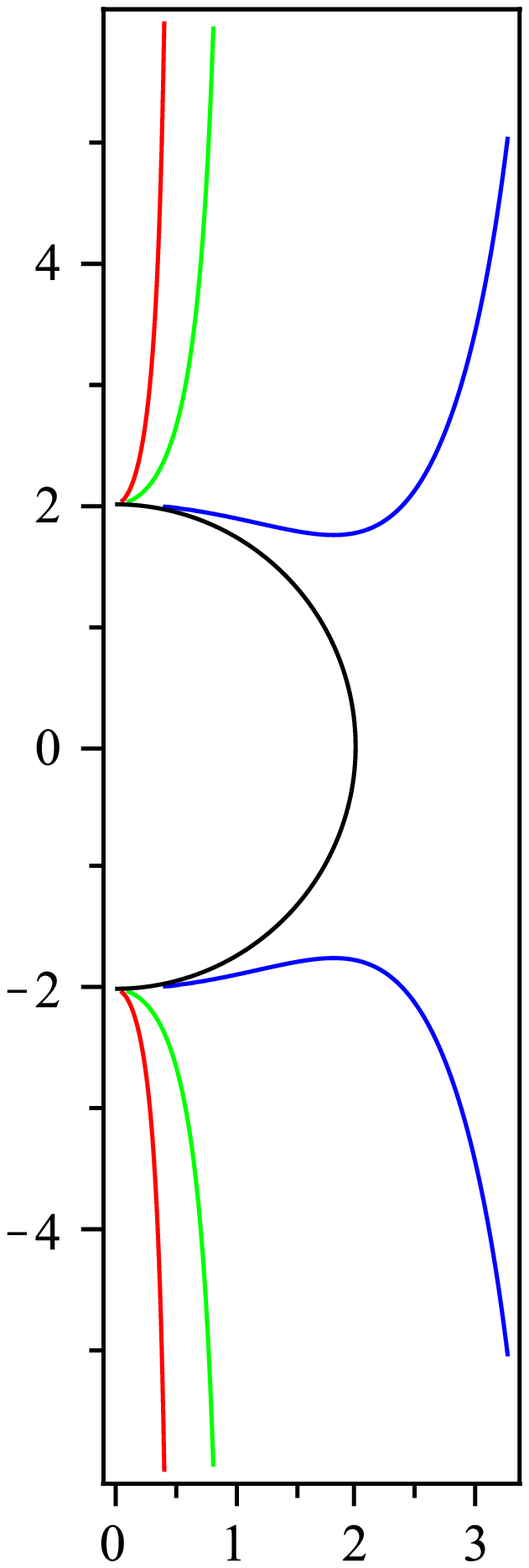} &
   \includegraphics[width=0.35\textwidth]{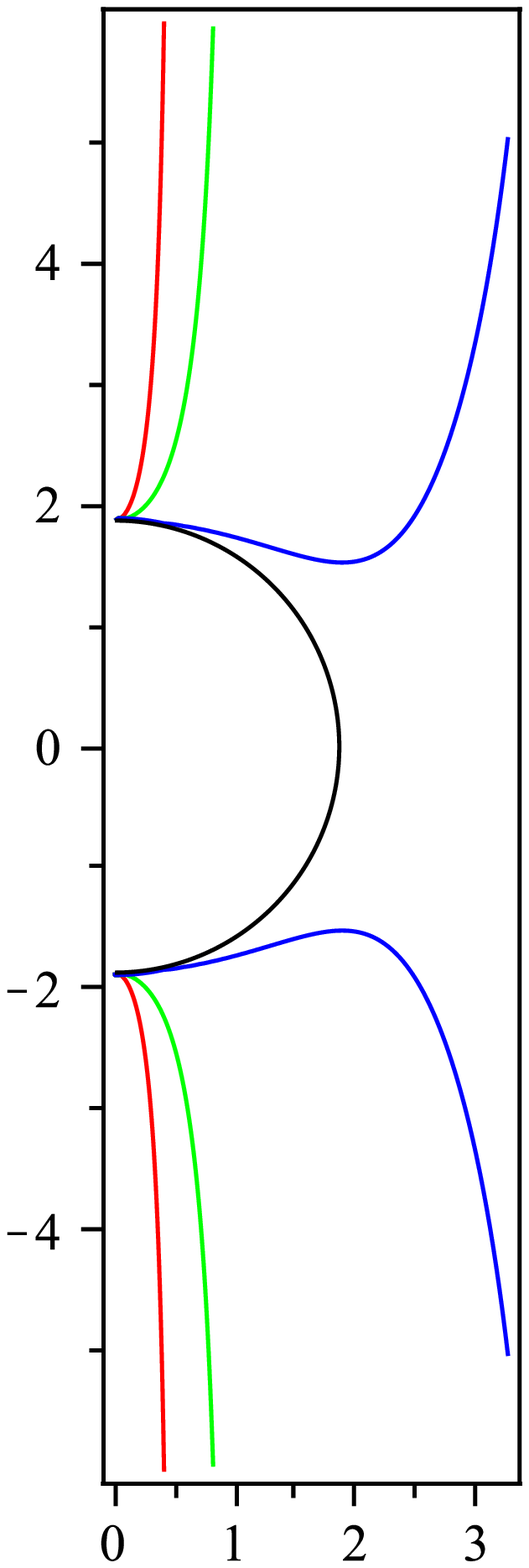} &
   \includegraphics[width=0.35\textwidth]{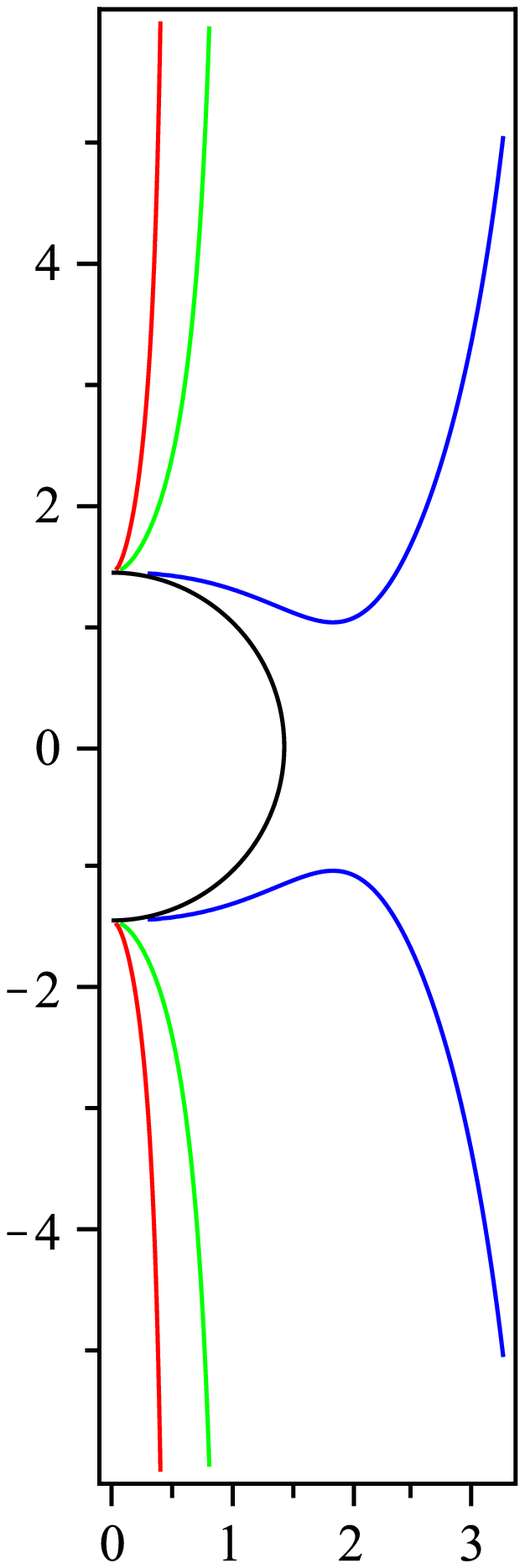}
 \end{tabular}
 \bigskip
 \bigskip
 \caption{{\footnotesize
 Similar plots as in previous figures depicting lines of constant $\xi$-values (\ref{Kerr_xi}), according to
 Example 1 for three different values of the Kerr's parameter $a$. Red lines correspond to $\xi=0.5$,
  green to $\xi=1$, and blue to $\xi=4$. Along these lines the electromagnetic field is, respectively,
  of the pure magnetic type, pure null type, and pure electric
  type, respectively, see (\ref{Inv1puro}). Green lines correspond to the light surfaces.}}
\end{figure*}

\begin{figure*}
 \centering
 \begin{tabular}{ccccc}
   $a=0$ & $a=0.5$  & $a=0.9$ \\
   \includegraphics[width=0.3\textwidth]{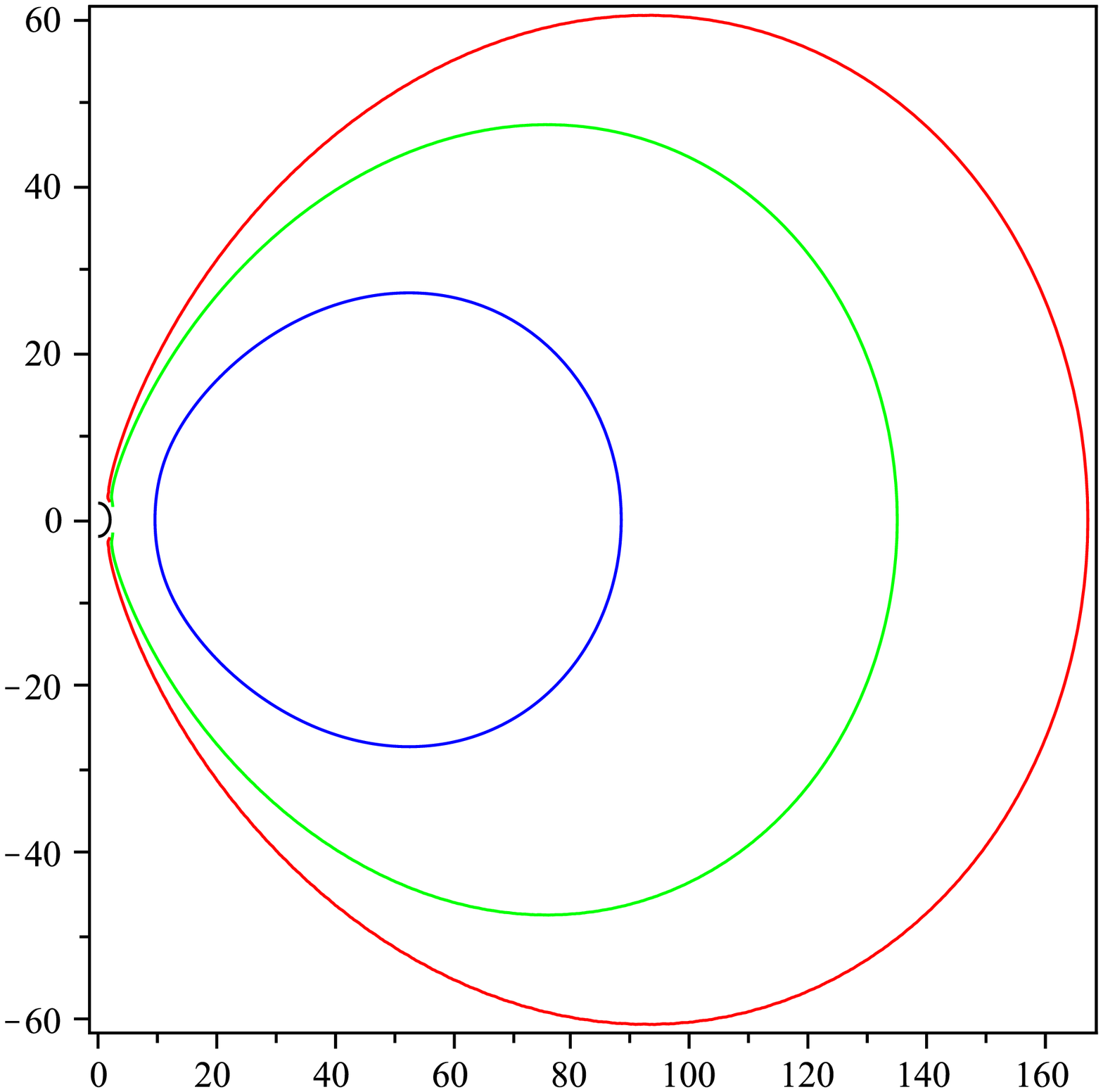} &
   \includegraphics[width=0.3\textwidth]{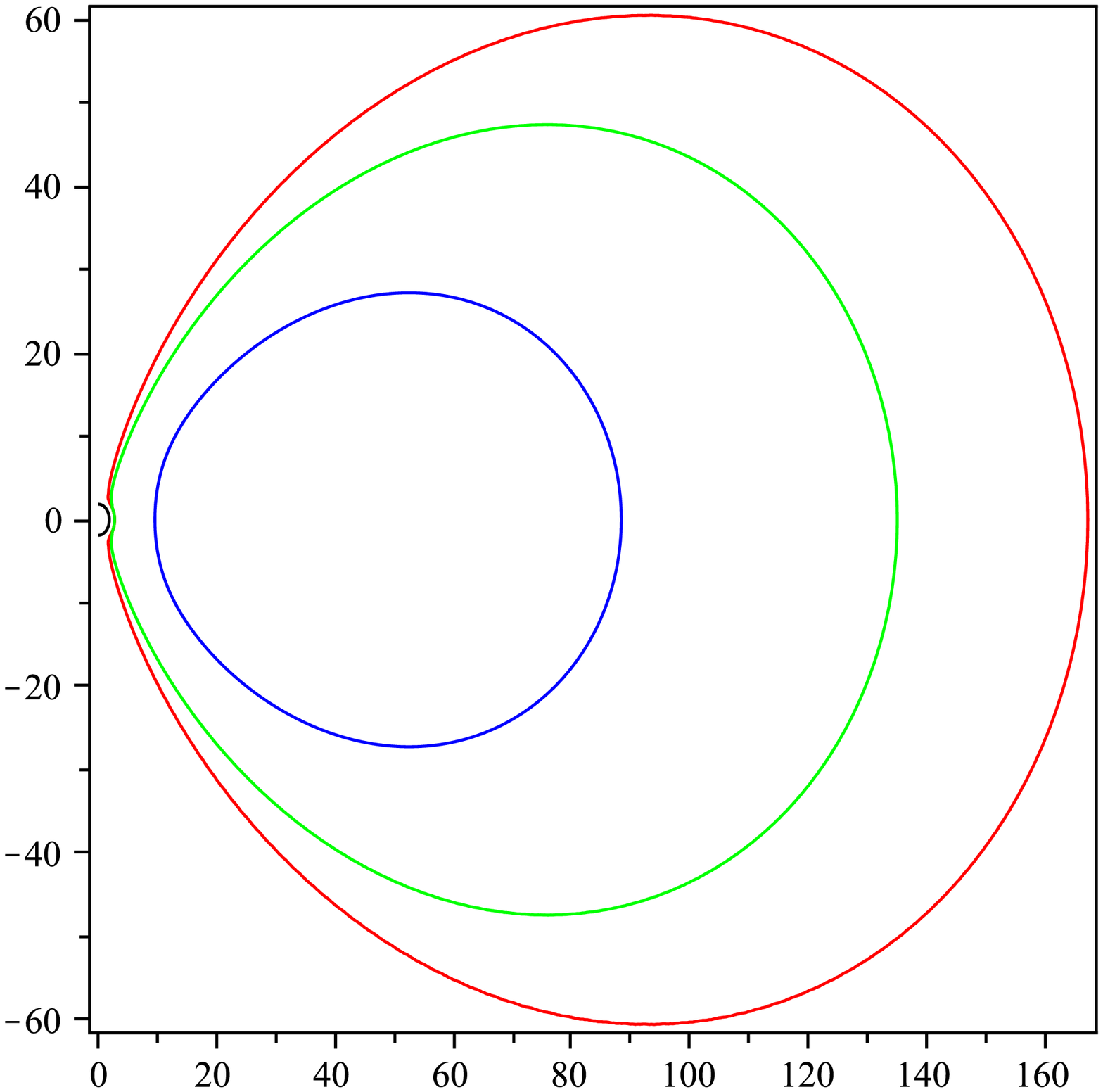} &
   \includegraphics[width=0.3\textwidth]{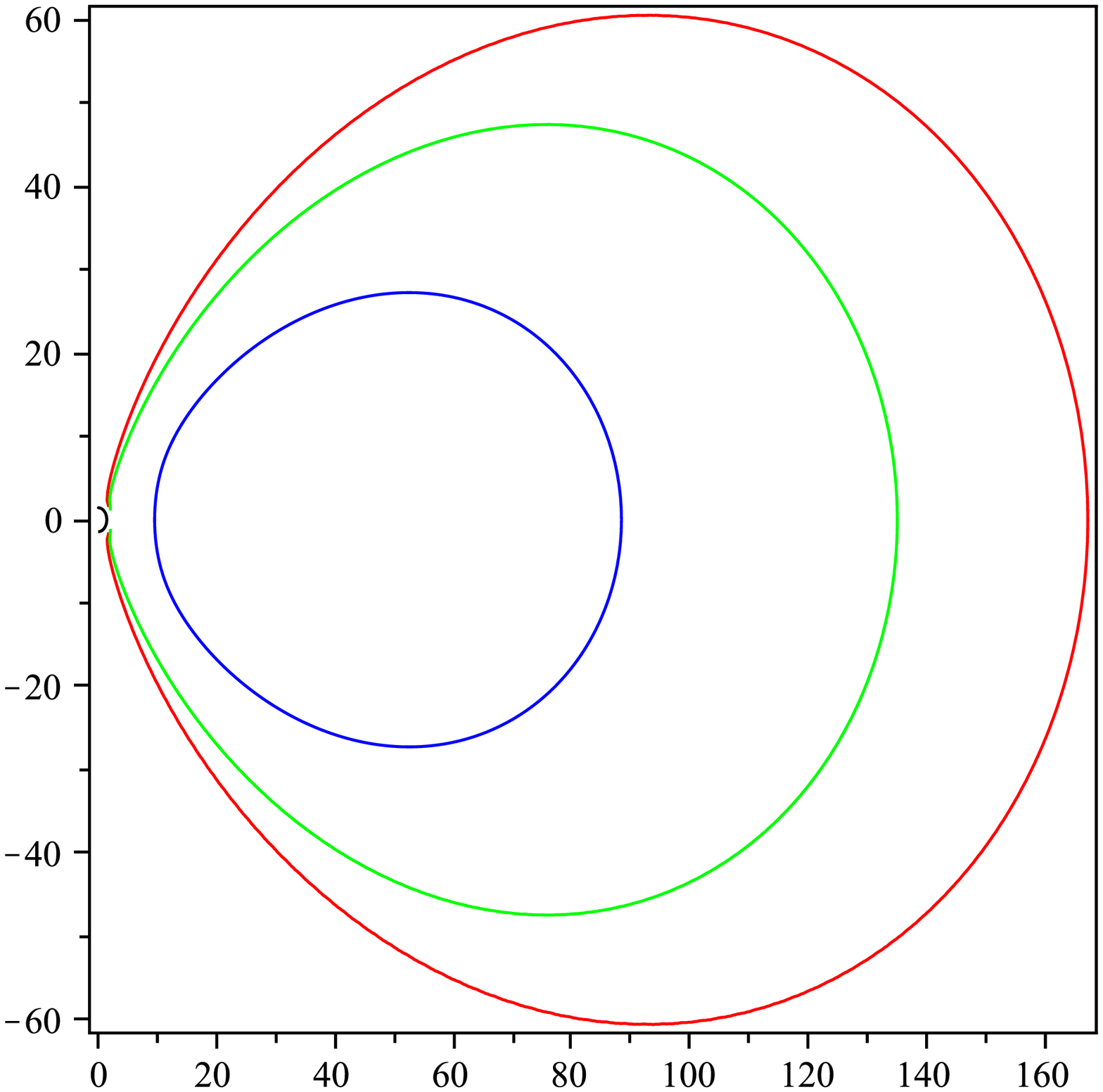}
 \end{tabular}
 \bigskip
 \bigskip
 \caption{{\footnotesize
 Similar contours plots as in the previous figure, but according to Example 2.
 Red lines correspond to $\xi=0.1$,
  green to $\xi=1$ and blue to $\xi=10$ on this lines the electromagnetic field is, respectively, of the pure magnetic type, pure null type and pure electric type, see (\ref{Inv1puro}). Green lines correspond to the light surfaces. }}
\end{figure*}

\subsection{The Kerr--Newman electromag\-ne\-tic field}
One example of a non-pure electromagnetic field is produced by a charged rotating black
hole. This field is not a test-one since it is a solution of the Einstein-Maxwell equations.
It has been already considered by Bini {\em et al.} \cite{Bini} and more recently, by Wylleman {\em et al.} \cite{WyllemanCQG}.
The latter group found that Carter observers, {i.e.}, those observers which measure a vanishing gravitational super-Poynting vector, also measure a
vanishing electromagnetic Poynting vector. Below we show that both the Poynting vector and the super-Poynting vector also vanish for the locally non-rotating observers as they approach the horizon, due to the frame dragging.

The Kerr--Newman solution describes the spacetime of a rotating black hole with charge
$Q$. It has the well-known form
$$
ds^2= \left(1-\frac{2mr-Q^2}{\rho^2}\right)dt^2+\frac{2a(2mr-Q^2)\sin^2\vartheta}{\rho^2} dt d\phi
-\frac{\rho^2}{\Delta_{KN}(r)}dr^2
$$
\begin{equation}
-\rho^2 d\vartheta^2
-\left(r^2+a^2+\frac{a^2(2mr-Q^2)\sin^2\vartheta}{\rho^2}\right)\sin^2 \vartheta\,d\phi^2~,
\end{equation}
where
\begin{equation}
\Delta_{KN}(r) = r^2-2Mr+a^2+Q^2,~~~~\rho^2=r^2+a^2\cos^2\vartheta~.
\end{equation}
For this metric, the horizons are located at $r_{\pm}=m\pm\sqrt{m^2-a^2-Q^2}$.

The electromagnetic four potential has the expression
\begin{equation}
A = -\frac{Qr}{\rho^2}\left(dt-a\sin^2\vartheta \,d\phi\right)~
\end{equation}
and the corresponding electromagnetic field is given by
$$
F=\frac{Q(r^2-a^2\cos^2\vartheta)}{\rho^4} dr\wedge\left( dt-a\sin^2\vartheta\,d\phi \right)+
$$
\begin{equation}
+\frac{2Qra\cos\vartheta}{\rho^4}\sin\vartheta d\vartheta\wedge\left[(r^2+a^2)d\phi-a\,dt\right]~.\label{FKerrNewman}
\end{equation}

In this spacetime we can introduce two new sets of observers, in the regions $0<r<r_-$ and $r>r_+$.
The first set is described by the monad field
$$
\tilde{\tau}= \frac{\sqrt{\Delta_{KN}(r)}}{\rho}\left( dt-a\sin^2\vartheta\,d\phi \right)
$$
\begin{equation}
=\frac{r^2+a^2}{\sqrt{(r^2+a^2)^2-\Delta_{KN} a^2\sin^2\vartheta}}
\left[\theta^{(0)}-\frac{a\sqrt{\Delta_{KN}}\sin\vartheta}{r^2+a^2}\theta^{(3)} \right]\label{Carter}
\end{equation}
together with the adapted tetrad to these observers
\begin{equation}
\begin{array}{lll}
\tilde{\theta}^{(0)}= \frac{\sqrt{\Delta_{KN}(r)}}{\rho}\left( dt-a\sin^2\vartheta\,d\phi \right), & ~ &
\tilde{\theta}^{(1)}= \frac{\rho}{\sqrt{\Delta_{KN}(r)}}dr,\\
\tilde{\theta}^{(2)}= \rho d\vartheta,&~&
\tilde{\theta^{(3)}}= \frac{\sin\vartheta}{\rho} \left[(r^2+a^2)d\phi-a\,dt\right]~.
\end{array}\label{cartertetrad}
\end{equation}
On the other hand, in the region between horizons, $r_-<r<r_+$, the monad field of the second set of observers is
\begin{equation}
\check{\tau}=\frac{\rho}{\sqrt{-\Delta_{KN}(r)}}dr~,\label{Carter2}
\end{equation}
with the adapted tetrad
\begin{equation}
\begin{array}{lll}
\check{\theta}^{(0)}= \frac{\sqrt{-\Delta_{KN}(r)}}{\rho}\left( dt-a\sin^2\vartheta\,d\phi \right), & ~ &
\check{\theta}^{(1)}= \frac{\rho}{\sqrt{-\Delta_{KN}(r)}}dr,\\
\check{\theta}^{(2)}= \rho d\vartheta,&~&
\check{\theta}^{(3)}= \frac{\sin\vartheta}{\rho} \left[(r^2+a^2)d\phi-a\,dt\right]~.
\end{array}\label{CarterNonRotatingtetrad}
\end{equation}
In this region, $\check{\theta}^{(1)}$ is a time-like covector, while $\check{\theta}^{(0)}$ is space-like one.
Notice that the monad field (\ref{Carter2})  is a non-rotating one $*(\check{\theta}^{(1)}\wedge d\check{\theta}^{(1)})=0$,
hence these Carter observers can be regarded also as locally non-rotating observers. Hence let us call
observers with four velocity field (\ref{Carter}), Carter observers, and observers with four velocity field (\ref{Carter2}),
locally non-rotating Carter observers.

In the regions $0<r<r_-$ and $r_+<r$, the observers represented by the velocity field (\ref{Carter}) measure
parallel electric and magnetic (co)vectors
\begin{equation}
\tilde{E}=\frac{Q(r^2-a^2\cos^2\vartheta)}{\rho^4}\tilde{\theta}^{(1)}~,
~~~~
\tilde{B}=\frac{2aQr\cos\vartheta}{\rho^4}\tilde{\theta}^{(1)}~.
\end{equation}
On the other hand, in region $r_-<r<r_+$, the observers with monad field (\ref{Carter2}) measure the parallel
electric and magnetic covectors
\begin{equation}
{\check E}=\frac{Q(r^2-a^2\cos^2\vartheta)}{\rho^4}{\check{\theta}}^{(0)},
~~~~
{\check B}=\frac{2aQr\cos\vartheta}{\rho^4}{\check{\theta}}^{(0)}.
\end{equation}
Hence, both sets of observers measure a vanishing Poynting vector.

From the above expressions, it is easy to see that the electromagnetic field is quite diverse \cite{Classif}.
All types of electromagnetic fields are present; for example, there is a pure electric type field on
the equatorial plane, a pure magnetic type on the surface
$r^2-a^2\cos^2\vartheta=0$, a null non-pure on the surface $r^4-6a^2r^2\cos^2\vartheta+a^4\cos^4\vartheta=0$, etc.

From (\ref{Carter}), in the regions $0<r<r_-$ and $r_+<r$, we see that the speed of
Carter observers (\ref{Carter}) as measured by the locally non-rotating observers is
\begin{equation}
{\bf \tilde{v}}=-\frac{a\sqrt{\Delta_{KN}}\sin\vartheta}{r^2+a^2}\theta^{(3)}~.
\end{equation}
Near the horizon the speed tends to zero (since $\Delta_{KN}\rightarrow 0$), meaning that both kind of observers become
comoving due to the dragging of reference frames.
In the region between both horizons, $r_-<r<r_+$, Carter observers and locally non-rotating observes become
totally comoving ones.

\subsubsection{Electromagnetic energy flux and superenergy flux in Kerr--Newman spacetime}
There is no local definition for the energy--momentum tensor of
the gravitational field in general relativity. Nevertheless, there is a tensor which could be used to describe locally
the strength of the gravitational field, the Bel-Robinson superenergy tensor \cite{Bel,Robinson}. It is constructed,
in a similar manner as the electromagnetic field, but using the Weyl curvature tensor:
\begin{equation}
T^{\alpha\beta\lambda\mu}={C^{\alpha\rho\lambda}}_{\sigma}{C^{\mu\sigma\beta}}_{\rho}+
C^{\alpha\rho\lambda} _{\;*~~~\sigma}C^{\mu\sigma\beta} _{\;*~~~\rho}~,
\end{equation}
where, the Hodge asterisk denote the dual conjugation on the pair of indexes above it,
\begin{equation}
C^{\alpha\rho\lambda} _{\;*~~~\sigma}=\frac{1}{2}{E^{\alpha\rho}} _{\beta\gamma}{C^{\beta\gamma\lambda}}_\sigma~.
\end{equation}
The Bel-Robinson tensor is symmetric in all its indices
\begin{equation}
T^{\alpha\beta\lambda\mu}=T^{(\alpha\beta\lambda\mu)}~ \label{SimetriaBR}
\end{equation}
and has properties similar to those of the traditional energy-momentum tensor. For instance, it possesses a
positive-definite timelike component, a causal ``momentum'' vector, its divergence vanishes (in vacuum), etc.
However, its units are the square of the energy density units.

The components of the Weyl tensor in the rotating tetrad basis (\ref{cartertetrad}) are
\begin{equation}
{\tilde C}_{(0)(1)(0)(1)}=-{\tilde C}_{(2)(3)(2)(3)}=2U(r,\vartheta)~,
\end{equation}
\begin{equation}
{\tilde C}_{(1)(2)(1)(2)}={\tilde C}_{(1)(3)(1)(3)}=-{\tilde C}_{(0)(2)(0)(2)}=-{\tilde C}_{(0)(3)(0)(3)}=U(r,\vartheta)~,
\end{equation}
\begin{equation}
{\tilde C}_{(0)(1)(2)(3)}=-2V(r,\vartheta)~,
\end{equation}
\begin{equation}
{\tilde C}_{(1)(2)(3)(0)}={\tilde C}_{(1)(3)(0)(2)}=-V(r,\vartheta)~,
\end{equation}
where
\begin{equation}
U(r,\vartheta)=\frac{mr^3-Q^2r^2-a^2(3mr-Q^2)\cos^2\vartheta}{(r^2+a^2\cos^2\vartheta)^2}~,
\end{equation}
and
\begin{equation}
V(r,\vartheta)=\frac{a(3mr^2-2 Q^2r-a^2m\cos^2\vartheta)\cos\vartheta}{(r^2+a^2\cos^2\vartheta)^2}~.
\end{equation}
The non-null components of the Bel-Robinson tensor in the tetrad basis (\ref{cartertetrad}) are
\begin{equation}
{\tilde T}^{(0)(0)(0)(0)}={\tilde T}^{(1)(1)(1)(1)}={\tilde T}^{(2)(2)(2)(2)}={\tilde T}^{(3)(3)(3)(3)}= 6C(r,\vartheta)\label{uno}~,
\end{equation}
\begin{equation}
{\tilde T}^{(2)(2)(0)(0)}={\tilde T}^{(3)(3)(0)(0)}=-{\tilde T}^{(1)(1)(2)(2)}=-{\tilde T}^{(1)(1)(3)(3)}= 4C(r,\vartheta)~,
\end{equation}
\begin{equation}
{\tilde T}^{(2)(2)(3)(3)}=-{\tilde T}^{(1)(1)(0)(0)}=2C(r,\vartheta)\label{tres}~,
\end{equation}
where
\begin{equation}
C(r,\vartheta)=\frac{m^2r^2-2mQ^2r+a^2m^2\cos^2\vartheta+Q^4}{(r^2+a^2\cos^2\vartheta)^4}~. \label{BelRobinsonCfunction}
\end{equation}
Notice that the two sets of tetrads (\ref{cartertetrad}) and (\ref{CarterNonRotatingtetrad}) are related thorough
\begin{equation}
\check{\theta}^{(0)}=i \tilde{\theta}^{(0)},~~~\check{\theta}^{(1)}=- i \tilde{\theta}^{(1)},~~~\check{\theta}^{(2)}=\tilde{\theta}^{(2)}, ~~~\check{\theta}^{(3)}=\tilde{\theta}^{(3)}~,
\end{equation}
hence, it is easy to obtain ${\check T}^{(\kappa)(\lambda)(\mu)(\nu)}$ from (\ref{uno})-(\ref{tres}).

The super-Poynting vector, representing the super-energy flux is defined as follows
\begin{equation}
{\cal P}_{(\alpha)}=b_{(\alpha)(\beta)}T^{(\beta)(\kappa)(\lambda)(\mu)}\tau_{(\kappa)}\tau_{(\lambda)}\tau_{(\mu)}~.
\end{equation}
Thus
\begin{itemize}
\item In the regions $r<r_-$ and $r>r_+$, using the monad $\tilde{\tau}={\tilde{\theta}}^{(0)}$, eq.~(\ref{Carter}), the super-Poynting vector vanishes
$$
{\tilde {\cal P}}_{(\alpha)}={\tilde b}_{(\alpha)(\beta)}{\tilde T}^{(\beta)(\kappa)(\lambda)(\mu)}{\tilde \tau}_{(\kappa)}
{\tilde \tau}_{(\lambda)}{\tilde \tau}_{(\mu)}=0~.
$$

\item In the region $r_-<r<r_+$, using the monad ${\check\tau}={\check \theta}^{(1)}$, eq.~(\ref{CarterNonRotatingtetrad}),  the super-Poynting vector vanishes
$$
{\check {\cal P}}_{(\alpha)}={\check b}_{(\alpha)(\beta)}{\check T}^{(\beta)(\kappa)(\lambda)(\mu)}{\check \tau}_{(\kappa)}
{\check \tau}_{(\lambda)}{\check \tau}_{(\mu)}=0~,
$$
hence rotating Carter observers and locally non-rotating Carter observers both measure a vanishing super Poynting vector.
\end{itemize}

The calculation of the super-Poynting vector for the locally non-rotating observers is more involved.
We use the non-rotating monad vector (\ref{MonadaNoRotante}) written in terms of the rotating tetrad.
We obtain
$$
{\cal P}=b_{(\alpha)(\beta)}T^{(\beta)(\kappa)(\lambda)(\mu)}\tau_{(\kappa)}\tau_{(\lambda)}\tau_{(\mu)}{\tilde\theta}^{(\alpha)}
$$
$$
=-18C(t,r)\tau_{(0)}\tau_{(3)}\left(\tau_{(0)} ^2 + \tau_{(3)} ^2 \right)
\left(\tau_{(3)} {\tilde \theta}^{(0)} + \tau_{(0)} {\tilde \theta}^{(3)} \right)
$$
\begin{equation}
= -18C(t,r)\tau_{(0)}\tau_{(3)}\left(\tau_{(0)} ^2 + \tau_{(3)}^2 \right)\theta^{(3)}~, \label{NonRotatingSuperpoynting}
\end{equation}
where $C(t,r)$ is defined in (\ref{BelRobinsonCfunction}). Locally non-rotating observers measure a net superenergy flux
in the $\theta^{(3)}$ direction in agreement with the result reported by Herrera {\em et al.} \cite{Herrera}. In addition,
they also suggest that this flux may be the cause of the reference frame dragging.

In order to deduce expression (\ref{NonRotatingSuperpoynting}), we have used the symmetry property of Bel-Robinson tensor (\ref{SimetriaBR}),
the shorthand notation $\tau=\tau_{(0)}{\tilde\theta}^{(0)}+\tau_{(3)}{\tilde\theta}^{(3)}$.
$$
\tau_{(0)}= \frac{r^2+a^2}{\sqrt{(r^2+a^2)^2-\Delta_{ KN} ^2\sin^2\vartheta}},
~~~~~
\tau_{(3)}=
\frac{a \sqrt{\Delta_{KN}}\,\sin\vartheta}{\sqrt{(r^2+a^2)^2-\Delta_{KN} ^2\sin^2\vartheta}}~,
$$
the property $\tau_{(0)} ^2 - \tau_{(3)} ^2 =1$. Also, $b_{(\mu)(\nu)}=g_{(\mu)(\nu)}-\tau_{(\mu)} \tau_{(\nu)} $
and
\begin{equation}
\theta^{(3)}=\frac{1}{\sqrt{(r^2+a^2)^2-\Delta_{KN} ^2~\sin^2\vartheta}}
\left[
a\sqrt{\Delta_{KN}} ~\sin\vartheta{\tilde\theta}^{(0)}+(r^2+a^2){\tilde\theta}^{(3)} \label{relacionTheta3}
\right]~.
\end{equation}
Notice that $\tau_{(3)}=0$ on the horizons, consequently ${\cal P}\rightarrow 0$ there. This does not
mean that there is no net super-energy flux at the horizon, but that non-rotating observers
become comoving with rotating Carter ones.

 As a final remark, it has to be noticed that Carter observers are not the only ones which
measure vanishing Poynting and super-Poynting vectors. In fact, as
Wylleman {\em et al.} have shown  \cite{WyllemanCQG}, any radially moving observer with respect to
the Carter observers also measures the vanishing of both Poynting vectors (These authors use the term
Weyl observers instead of the Carter observer used here).

In the regions $r<r_-$ and $r>r_+$ these {\it Wylleman observers} have the following 4-velocity
\begin{equation}
\breve{\tau}=\cosh(\Psi)\tilde{\theta}^{(0)}-\sinh(\Psi)\tilde{\theta^{(1)}},
\end{equation}
where $\Psi=\Psi(x)$ is an arbitrary function of the space-time coordinates.

These observers can readily be extended to the region $r_-<r<r_+$; their 4-velocity is
\begin{equation}
\bar{\tau}=\cosh(\Upsilon)\tilde{\theta}^{(1)}-\sinh(\Upsilon)\tilde{\theta^{(0)}}~,
\end{equation}
where $\Upsilon=\Upsilon(x)$ is also an arbitrary function of the coordinates. They move in the {\it time} direction, that of the spatial tetrad vector
$\tilde{X}_{(0)}=\partial_t$, without detecting Poynting and super Poynting flux vectors.

Contrary to our previous ideas,  it seems that the state of motion of
both the electromagnetic field and the gravitational-superenergy field cannot be established
using the criterion of the cancellation of the Poynting vector or the Poynting supervector, and
it is not possible to speak about their state of motion at all.

\section{Vanishing Poynting observers in general relativity}
As the last example shows, observers measuring a null Poynting vector are not unique. In a recent work by Wylleman et al. \cite{WyllemanCQG}
it is shown how to find all these observers.  Now, using the presented formalism, and without considering any concrete space-time, we can generalize
our previous results to show that these observers can always be introduced for any non-null electromagnetic field.

In order to do so, we introduce an arbitrary unitary time-like covector $\tau$, and an arbitrary space-like covector $p$, together with
\begin{equation}
\hat{p}=\frac{1}{\sqrt{-\,p\cdot p}} ~ p~.
\end{equation}

\begin{enumerate}
\item For the pure-electric type case, the electromagnetic field tensor is always a simple bivector of the form $F_E=p\wedge\tau$ [compare with (\ref{FIIa})];
      observers measuring a vanishing magnetic field have the following four-velocity field:
      \begin{equation}
      \tau_E=\cosh(\psi)\tau-\sinh(\psi)\hat{p}~.
      \end{equation}
      One can check that $B=*(\tau_E\wedge F_E)\equiv0$.
\item For the pure-magnetic type case, the electromagnetic field tensor is always the dual conjugate of a simple bivector of the form
      $F_B=*(p\wedge\tau)$ [compare with (\ref{FIa})];
      observers measuring a vanishing electric field have the following four-velocity field:
      \begin{equation}
      \tau_B=\cosh(\psi)\tau-\sinh(\psi)\hat{p}~.
      \end{equation}
      One can check that $E=*(\tau_B\wedge F_B)\equiv0$.
\item The electromagnetic field tensor of a non-pure electromagnetic field can always be written as
      $F_{e}=\cos(\alpha){\cal F}_e-\sin(\alpha)*{\cal F}_e$, see Section 6 and (\ref{impuro2}). For the electric type case, ${\cal F}_e$ is a simple
      bivector of the form ${\cal F}_e=p\wedge \tau$ [compare with (\ref{impuro3})]; observers measuring parallel electric and magnetic fields
      have the following four-velocity field:
      \begin{equation}
      \tau_e=\cosh(\psi)\tau-\sinh(\psi)\hat{p}~.
      \end{equation}
      One can check that
      \begin{equation}
      E_e =*(\tau_e\wedge *F_e)=\cos\alpha *(\tau_e\wedge*{\cal F}_e)~,
      \end{equation}
      \begin{equation}
      B_e =*(\tau_e\wedge F_e)=-\sin\alpha *(\tau_e\wedge*{\cal F}_e)~,
      \end{equation}
      are proportional to each other.

\item The electromagnetic field tensor of a non-pure electromagnetic field can always be written in the form
      $F_{b}=\cos(\alpha){\cal F}_b-\sin(\alpha)*{\cal F}_b$, see Section 6, and (\ref{impuro2}). For the magnetic type case, ${\cal F}_b$ is of the
      form ${\cal F}_b=*(p\wedge \tau)$ [compare with (\ref{impuro4})]; observers measuring parallel electric and magnetic fields
      have the following four-velocity field:
      \begin{equation}
      \tau_b=\cosh(\psi)\tau-\sinh(\psi)\hat{p}~.
      \end{equation}
      One can check that
      \begin{equation}
      E_b= *(\tau_b\wedge *F_b)=-\sin\alpha *(\tau_b\wedge{\cal F}_b)~,
      \end{equation}
      \begin{equation}
      B_b= *(\tau_b\wedge *F_b)=-\sin\alpha *(\tau_b\wedge{\cal F}_b)~,
      \end{equation}
      are proportional to each other.
\end{enumerate}
 Here, $\psi=\psi(x)$ is an arbitrary function of space-time coordinates. In all cases, the new observers
 move with a speed $v=\tanh(\psi)$ in the direction of the $p$-vector with respect to a family of observers
 with 4-velocity field $\tau$.

Thus, contrary to our previous ideas in \cite{Nuestroarticulo}, these results show that the state of motion of
the electromagnetic field cannot be established using the criterion of the cancellation of the Poynting vector,
since there is a multitude of observers in relative motion to each other that measure a vanishing Poynting vector.

\section{Conclusions}

In this work, we have extended our previous study on rotating electromagnetic fields in special relativity \cite{Nuestroarticulo} by
considering axially symmetric test fields outside the event horizons of Schwarzschild and Kerr black holes. We have also considered
the non-test electromagnetic field of the Kerr-Newman black hole. These rotating fields are also characterized by having an angular momentum density
regarding some reference frame. Similarly to the previous work, we have introduced two families of observers, locally
non-rotating observers and rotating vanishing Poynting observers.

The reference frame of the locally non-rotating observers was used to describe the electric and magnetic fields, and also the
charge and current densities associated with the different fields considered here; as well as to characterize
the motion of the vanishing-Poynting observers. In their turn,
the vanishing Poynting observers were used to describe, in their frame of reference, the electric and magnetic fields.
Considering the nature of the electromagnetic field, it was necessary to divide the consideration of this last family of
observers into two cases:
\begin{enumerate}
\item If the electromagnetic field is of the pure type (the second electromagnetic invariant vanishes) we introduce
observers for which either the electric field cancels or the magnetic field cancels. These observers were labeled with the
subscripts I and II.

\item If the electromagnetic field is of the non-pure type (the second invariant does not cancel) we introduce
observers for which the electric and magnetic fields are mutually parallel.These observers were labeled with the
subscripts A and B.
\end{enumerate}

Due to the presence of light surfaces, regions where the electromagnetic field is of the pure null type,
it was necessary to divide the observers in each of the previous cases into two subfamilies,
each one on each side of the light surface.

In the case of pure fields, in the locally non rotating reference frame, for each electromagnetic field of the form (\ref{Fourpotential}), with
 $M=M(N)$ ($I_2=0$ for this choice), two subfamilies of rotating observers measuring a vanishing Poynting vector were introduced:
 In the pure magnetic type region, observers $I$ with a
velocity field given by (\ref{vI}), while in the pure electric type region, observers $II$ with a velocity field given by (\ref{vII}) both fields are
related to the function $\xi$ defined in (\ref{Kerr_xi}).
Also, in the locally non-rotating reference frame, since the magnetic field does not depend on the function $M$, for a given magnetic field (\ref{Bpure_non_rotating frame}), there were associated several different unipolar induced electric fields (\ref{Epure_non_rotating frame}), several different charge densities (\ref{Qdensity_no_rotating}) and several different current densities (\ref{Jpure_no_rotating}), one for each $M$.
An important difference with respect to our previous work \cite{Nuestroarticulo}
is that there is a contribution to the unipolar induced electric field from the
frame dragging. As a matter of fact, frame dragging alone is enough to induce an electric field.

 To show these latter features, several examples have been given in Schwarz\-schild and Kerr spacetimes,
 different dipolar-like magnetic fields (vanishing at the horizon) were used to show the complex structure of
 the charge and the current densities. To see the nature of the electromagnetic field and visualize the light surface,
 we have also provided some plots showing the profiles of revolution of the first electromagnetic field invariant.

 We have also shown that in the special case when the pure electromagnetic field is of the pure magnetic type everywhere outside the horizon,
 ideal magnetohydrodynamic conditions are satisfied. Consequently, these fields might model the magnetospheres of neutron stars,
 or the final evolution stage of accretion disks around non rotating black holes, once the infalling matter supply has been depleted.

In the case of non-pure fields, instead of test fields, we have considered the electromagnetic field of the
Kerr-Newman solution. We have found that there are regions where the electromagnetic field is also pure, as some
other authors have pointed out \cite{Classif,Bini}.
Using the locally non-rotating Wylleman observers, we have found that the electromagnetic Poynting vector and the gravitational
super-Poynting vector vanish for these observers in the region between horizons. These observers also measure parallel electric and magnetic
vectors in the non-pure regions, confirming results in \cite{WyllemanCQG}. Also, in this example, we found how
the rotating Carter observers and locally rotating observers become comoving ones in the region between the horizons, due to the effect of
frame dragging.

Although we have restricted our considerations to stationary electromagnetic fields in the Kerr and Kerr-Newman spacetimes, in section 8, we extended
our methodology to obtain the vanishing Poynting observers for non null electromagnetic fields in other geometries, using only the algebraic structure of the electromagnetic field tensor.

The examples presented here can be used not only for pedagogical purposes, but also to complement studies about numerical \cite{McKinney2006} and exact \cite{Lupsasca2015,Menon2007} solutions of black hole magnetospheres or even unipolar induction \cite{Okamoto2012,Lyutikov2011,Toma2014,Gralla2016}.
In particular, these examples suggest that black hole magnetospheres may be quite diverse.

\section*{Data Availability}
No data were used to support this study.

\section*{Conflicts of interest}
The authors declare that they have no conflicts of interest.

\section*{Acknowledgement}
We appreciate the kind comments on this work of L. Wylleman,
L. Filipe O Costa, and J. Nat\'ario.
M.A. Mu\~niz Torres appreciates the support provided by the Consejo
Nacional de Ciencia y Tecnolog\'ia (CONACyT).
This work was funded by the University of Guadalajara through PROSNI program.

\end{document}